\documentclass[letter,reqno]{qt}
\title{
Reducibility Theory and Ergodic Theorems\\
for Ergodic Quantum Processes
}
\date{} 

\usepackage{authblk}
\author[1, 2]{Owen Ekblad\orcidlink{0009-0006-0834-0327}\thanks{ekbladow@msu.edu}}
\author[1]{Jeffrey Schenker\orcidlink{0000-0002-1171-7977}\thanks{schenke6@msu.edu}}
\affil[1]{Department of Mathematics\\
Michigan State University}
\affil[2]{Department of Mathematics and Center for Quantum Mathematics and Physics\\
University of California, Davis}

\begin{document}
\pagenumbering{arabic}
\lhead{\thepage}
\maketitle

\begin{abstract}
We develop a Perron-Frobenius type theory for products of random quantum channels acting on finite-dimensional matrix algebras sampled from a stationary and ergodic stochastic process, which we term ergodic quantum processes.  
This serves as a unifying framework for many models, including i.i.d., Markovian, periodic, and quasi-periodic models.
We establish various characterizations of irreducibility, from which we recover a number of general ergodic theorems.
We then analyze some specific examples, and, in particular, give a refinement of our theory in the i.i.d. case. 
\end{abstract}

\section{Introduction}\label{Sec:Introduction}
%
%
%
%
%
%
%
%
%
%
%
%
%
%
%
%
%
During the course of analysis of quantum systems, there are many situations where one encounters repeated compositions of quantum channels, which, for the purposes of this work, we define to be completely positive and trace-preserving linear maps $\psi:\matrices\to\matrices$, where $\matrices$ is the set of $d\times d$ matrices with entries in $\mbC$. 
For example, such compositions describe general open quantum system dynamics under the repeated interactions formalism \cite{Bruneau2014RepeatedSystems, Ciccarello2022QuantumInteractions}. 
For another example, when one studies matrix product states, such compositions show up naturally during the course of analyzing two-point correlation functions on quantum spin chains \cite{Fannes1992FinitelyChains}. 
Historically, most of the analysis of repeated compositions of quantum channels has been restricted to the homogeneous case, i.e., studying asymptotic behavior of a single quantum channel when repeatedly composed with itself. 
This theory is well-developed and can be studied using Perron-Frobenius type methods analogous to standard analysis of Markov chains on finite state spaces. 
There are, however, many physically-natural situations wherein one is motivated to understand general compositions of the form 
\begin{equation*}
    \phi_n\circ\cdots\circ\phi_1
\end{equation*}
where $\{\phi_1, \dots, \phi_n\}$ is a collection of distinct quantum channels.
In this work, we analyze this more general situation under the assumption that $\{\phi_1, \dots, \phi_n\}$ is sampled from a stationary and ergodic stochastic process.
Many authors have considered such compositions under certain stochastic assumptions, such as i.i.d. or Markovian assumptions \cite{Bruneau2008RandomSystems, Bougron2022MarkovianSystems}, or other assumptions, such as the so-called eventually strictly positive regime considered in \cite{Movassagh2022AnStates}.
However, to the authors' knowledge, none have considered the fully general setting we consider here. 
This work establishes a general Perron-Frobenius type framework with which one can understand a very broad class of compositions of quantum channels sampled from stationary stochastic processes, including each of the references we have already mentioned. 
To most conveniently describe our work, we require a small amount of notation. 
Let $\channels$ denote the set of quantum channels on $\matrices$, i.e., the set of linear maps $\psi:\matrices\to\matrices$ that are both completely positive and trace-preserving. 
We also fix $(\Omega, \mcF, \mu)$ a probability space, and take $\phi:\Omega\to\channels$ to be a measurable quantum channel-valued map\footnote{We endow $\channels$ with the Borel $\sigma$-algebra induced by the operator norm on the set of linear maps $\psi:\matrices\to\matrices$.}, where we write $\phi_\omega$ to denote the evaluation of $\phi$ at $\omega\in\Omega$. 
Let $\theta:\Omega\to\Omega$ be an invertible, ergodic, and measure-preserving transformation, by which we mean that $\mu[\theta^{-1}(A)] = \mu[A]$ for all $A\in\mcF$ (measure-preservation) and $\theta^{-1}(A) = A$ implies $\mu[A]\in\{0, 1\}$ for all $A\in\mcF$ (ergodicity).
Then the precise mathematical objects we are interested in understanding are the compositions
\begin{equation*}
    \phi_{\theta^n(\omega)}\circ\cdots\circ\phi_{\theta(\omega)}
\end{equation*}
where $\omega\in\Omega$ is arbitrary. 
For all $n\in\mbN$, we define $\phin:\Omega\to\channels$ to be the above composition, i.e., $\phin_\omega :=  \phi_{\theta^n(\omega)}\circ\cdots\circ\phi_{\theta(\omega)}$.
We call the tuple $(\theta, \phi)$ the \textit{ergodic quantum process} defined by $\theta$ and $\phi$.\footnote{
It is worth noting that if we define $\phi_n:\Omega\to\channels$ by $\phi_{n; \omega} = \phi_{\theta^n(\omega)}$ for $n\in\mbZ$, the stochastic process $(\phi_n)_{n\in\mbZ}$ is stationary in the sense that 
\begin{equation*}
    \operatorname{Law}\seq{\phi_{i_1}, \dots, \phi_{i_k}}
    =
    \operatorname{Law}\seq{
    \phi_{i_1 + m}, \dots, \phi_{i_k + m}
    }
\end{equation*}
for all $\set{i_j}_{j=1}^k\subset\mbZ$ and $m\in\mbZ$. 
Moreover, given such a stochastic process $(\phi_n)_{n\in\mbZ}$, there is a probability space $(\Omega, \mcF, \mu)$, a measure-preserving transformation $\theta:\Omega\to\Omega$, and a random quantum channel $\phi':\Omega\to\channels$ such that $(\phi_n)_{n\in\mbZ}$ is naturally obtained as the  quantum process defined by $(\theta, \phi')$; see \cite[Ch. 25]{Kallenberg2021FoundationsProbability} for more details. In 
}
Our analysis of ergodic quantum processes draws inspiration from the Perron-Frobenius theory of matrices with positive entries \cite{Perron1907ZurMatrices, Frobenius1912UberElementen}, later generalized to positive maps on finite-dimensional $C^*$-algebras \cite{Evans1978SpectralC-Algebras}.
There, the reducibility theory is most naturally posed in terms of projections $p\in\matrices$, where we call $p\in\matrices$ a projection if $p = p^2 = p^*$ where $(\cdot)^*$ denotes conjugate transpose.
Then in the framework of \cite{Evans1978SpectralC-Algebras}, $\psi\in\channels$ is said to be reduced by a projection $p\in\matrices$ if 
\begin{equation*}
    \psi(p\matrices p)
    \subseteq 
    p\matrices p,
\end{equation*}
and $\psi$ is called irreducible if the only nonzero $p$ is $p = \mbI$, where $\mbI\in\matrices$ is the identity. 
Here, the natural reducibility theory is formulated in terms of \textit{random} projections, where, given a random matrix $p:\Omega\to\matrices$ with $p$ a projection $\mu$-almost surely, we say that $p$ reduces the ergodic quantum process $(\theta, \phi)$ if 
\begin{equation*}
    \phi_{\theta(\omega)}
    \seq{
    p_\omega\matrices p_\omega
    }
    \subseteq
    p_{\theta(\omega)}\matrices p_{\theta(\omega)}
\end{equation*}
holds for $\mu$-almost every $\omega$.
We call such $p$ a reducing projection for $(\theta, \phi)$, and we say $p$ is nontrivial if $\mu\brackets{p \neq 0} > 0$.
We write $\redproj$ to denote the set of nontrivial reducing projections of $(\theta, \phi)$.
Note that $\redproj\neq\emptyset$ for any $(\theta, \phi)$, since $\mbI\in\redproj$.
Moreover, $\redproj$ is ordered with respect to the cone of positive semidefinite matrices in $\matrices$, i.e., we write $p\leq q$ for two random projections $p, q:\Omega\to\matrices$ if $q_\omega - p_\omega$ is positive semidefinite almost surely. 
Our first result gives the existence of minimal elements with respect to this ordering and classifies such minimal elements. 
Let $\mbP_d$ denote the set of positive semidefinite matrices, let $\states$ denote the set of $\arbrho\in\mbP_d$ with $\operatorname{Tr}(\arbrho) = 1$. 
Define $\rmatrices$ to be the $\mbC$-vector space of measurable functions $x:\Omega\to\matrices$ up to almost sure equivalence, and define $\mbP_d^\Omega\subset\rmatrices$ and $\rstates\subset\rmatrices$ analogously. 
\begin{restatable}[]{thm}{Minimalprojectionstheorem}\label{Thm:Minimal projections}
 Let $(\theta, \phi)$ be an ergodic quantum process.
 Then there exists a minimal element $p$ in $\redproj$, i.e., there is $p\in\redproj$ such that $q\leq p$ implies $q = p$ almost everywhere for all $q\in\redproj$. 
 Moreover, the following are equivalent. 
 \begin{enumerate}[label=(\alph*)]
    \item $p\in\redproj$ is minimal.

    \item There is a unique $\uniqrho\in\rstates$ such that $\operatorname{ran}(p_\omega) = \operatorname{ran}(\uniqrho_\omega)$ and $ \phi_{\theta(\omega)}(\uniqrho_\omega)
            =
        \uniqrho_{\theta(\omega)}$
    almost surely.

    \item The $\mbC$-vector space
    \begin{equation*}
        \operatorname{Fix}_p
            :=
        \set{
        x\in\rmatrices
            \,\,:\,\,
        p_\omega x_\omega p_\omega = x_\omega
        \text{ and }
        \phi_{\theta(\omega)}(x_\omega)
        =
        x_{\theta(\omega)}\text{ almost surely}
        }
    \end{equation*}
    is 1-dimensional, with $\operatorname{Fix}_p = \mbC\uniqrho$ for some $\uniqrho\in\rstates$ such that $\operatorname{ran}(p_\omega) = \operatorname{ran}(\uniqrho_\omega)$ almost surely.
 
    \item For all $\arbrho\in\rstates$ and  $x\in\mbP_d^\Omega$ with $\mu[x\neq 0]>0$, if $p\arbrho p = \arbrho$ and $pxp = x$, then
         \begin{equation*}
         \mu\left[
         \omega\in\Omega\,\,:\,\,
         \text{there exists $n\in\mbN$ such that }
                \tr{
                \phin_\omega(\arbrho_\omega)
                x_{\theta^n(\omega)}
                }>0
         \right] 
         =1.
        \end{equation*}
    \end{enumerate}
\end{restatable}
In the case that $\redproj = \set{\mbI}$, we call $(\theta, \phi)$ irreducible. 
More generally, if $\redproj$ contains only one element, we call $(\theta, \phi)$ dynamically ergodic. 
As we can see from the above, dynamical ergodicity of $(\theta, \phi)$ is equivalent to the uniqueness of $\uniqrho\in\rstates$ such that 
\begin{equation*}
    \phi_{\theta(\omega)}(\uniqrho_\omega)
    =
    \uniqrho_{\theta(\omega)}
\end{equation*}
holds for almost every $\omega\in\Omega$.
When $(\theta, \phi)$ is dynamically ergodic, we call such $\uniqrho$ the unique stationary state for $(\theta, \phi)$.
Our next theorem deals with this dynamically ergodic case: 
if $(\theta, \phi)$ is dynamically ergodic with unique stationary state $\uniqrho$, we show that, given any initial state $\arbrho$ whose range projection is subordinate to $\uniqrho$, time averages of quantum expectations of an observables $x$ with respect to $\arbrho$ converge to the disorder average of the quantum expectation of $x$ with respect to $\uniqrho$. 
We let $\mbE_\mu[\cdot]$ denote $\int_\Omega (\cdot)\,\dee\mu$.
\begin{restatable}[]{thm}{Dynamicallyergodic}\label{Thm:Dynamically ergodic theorem}
Let $(\theta, \phi)$ be an ergodic quantum process, fix minimal $p\in\redproj$, and let $\uniqrho:\Omega\to\states$ be as in Theorem \ref{Thm:Minimal projections} (b).
Then for all $\arbrho\in\rstates$ and essentially bounded $x\in\rmatrices$, if $p\arbrho p = \arbrho$ and $pxp = x$, then
\begin{equation*}\label{Eqn:Thm:Dynamically ergodic theorem, Eqn 1}
    \lim_{N\to\infty}
    \frac{1}{N}
    \sum_{n=1}^N
    \tr{
    \phin_\omega(\arbrho_\omega)
    x_{\theta^n(\omega)}
    }
    =
    \mbE_\mu[\tr{\uniqrho x}]
\end{equation*}
holds for $\mu$-almost every $\omega\in\Omega$. 
If, moreover, $(\theta, \phi)$ is dynamically ergodic with unique stationary state $\uniqrho\in\rstates$, then any $\arbrho\in\rstates$ satisfies 
\begin{equation*}
    \lim_{N\to\infty}
        \frac{1}{N}
        \sum_{n=1}^N
       \phin_\omega(\arbrho_\omega)
        =
        \mbE_\mu[\uniqrho]
\end{equation*}
for $\mu$-almost every $\omega$. 
\end{restatable}
As we show below, for any ergodic quantum process $(\theta, \phi)$, there is $\recP\in\redproj$ such that for any $\arbrho\in\rstates$ with 
\begin{equation*}
    \phi_{\theta(\omega)}(\arbrho_\omega)
    =
    \arbrho_{\theta(\omega)}
\end{equation*}
almost surely, we have that $\proj{\arbrho}\leq \recP$, where $\proj{\arbrho}\in\rmatrices$ denotes the (random) range projection of $\arbrho$.
We call $\recP$ the recurrent projection, and we call $\transP:= \mbI - \recP$ the transient projection. 
The last piece of the reducibility theory we develop here is the following decomposition theorem. 
\begin{restatable}[]{thm}{Recurrencetheorem}\label{Thm:Recurrence}
Let $(\theta, \phi)$ be an ergodic quantum process with recurrent projection $\recP$ and transient projection $\transP$. 
Then there is a finite set $\set{p_i}_{i=1}^n$ of minimal elements in $\redproj$ with $p_i p_j = p_j p_i = 0$ whenever $i\neq j$ and 
\begin{equation*}
    \sum_{i=1}^n p_i = \recP. 
\end{equation*}
Furthermore, any $\arbrho\in\rstates$ satisfies
\begin{equation*}
    \lim_{N\to\infty}
    \frac{1}{N}
    \sum_{n=1}^{N}
    \int_{\Omega}
    \tr{\phin_\omega (\arbrho_\omega)
        \transP_{\theta^n(\omega)}
    }\,\dee\mu(\omega)
    =
    0.
\end{equation*}
\end{restatable}
\begin{remark}
    The set $\set{p_i}$ need not be unique, which is clearly seen by noting that the identity quantum channel $\operatorname{Id}:\matrices\to\matrices$ has recurrent projection $\recP = \mbI$ and minimal reducing projections are simply rank one projections, but $\mbI = \sum_{i=1}^d \ketbra{\psi_i}{\psi_i}$ holds for any orthonormal basis $\set{\ket{\psi_i}}_{i=1}^d$ of $\mbC^d$.  
\end{remark}
\subsection{Physical motivation}
As mentioned in the above, our work is motivated on two fronts: namely, by both open quantum dynamics and by quantum spin chains. 
We now describe both motivations in further detail. 
\subsubsection{Open quantum dynamics}
It has long been understood that the description of a quantum system without explicitly including the environment requires the density operator formulation of quantum mechanics \cite{Neumann1927WahrscheinlichkeitstheoretischerQuantenmechanik, Neumann1932MathematischeQuantenmechanik}. 
A quantum mechanical system $\mcS$ interacting with an external environment $\mcE$ (which is sometimes called a \textit{bath}) is called open.
Such systems are ubiquitous: 
environmental interactions are often unavoidable in physical systems, and in some cases, system-bath interactions may even be engineered to accomplish tasks useful in applications \cite{Verstraete2009QuantumDissipation}. 
Under the assumption that $\mcS$ and $\mcE$ are only weakly coupled---by which we mean that the interaction of $\mcS$ with $\mcE$ only weakly affects the state of $\mcE$---the dynamics induced on $\mcS$ are well-approximated by the following discrete-time Markovian process: 
if the state of $\mcS$ is described by the density matrix $\arbrho_0$ at time $t = 0$, and the state of $\mcE$ is described by the density matrix $\rho$ at time $t = 0$, then the state $\arbrho_1$ of $\mcS$ at time $t = 1$ is given by
\begin{equation*}
    \arbrho_1
        =
    \operatorname{Tr}_{\mathcal{E}}\big( U(\arbrho_0\otimes \rho)U^*\big),
\end{equation*}
where $\operatorname{Tr}_{\mathcal{E}}\seq{\cdot}$ denotes the partial trace and $U$ describes the unitary evolution of the joint system $\mcS\otimes\mcE$ after one time step.
The assumption of weak interactions between $\mcS$ and $\mcE$ is realized by taking the state of $\mcE$ at time $t = 1$ to be described by $\rho$, i.e., that the interaction between $\mcS$ and $\mcE$ has little effect on the state of $\mcE$ itself.
By letting $\phi$ denote the map $\arbrho\mapsto \operatorname{Tr}_{\mathcal{E}}\big( U(\arbrho\otimes \rho)U^*\big)$, we see that this open quantum dynamical system is given by 
\begin{equation*}
    \arbrho_n = \phi^n(\arbrho_0),
\end{equation*}
where $\arbrho_n$ denotes the state of $\mcS$ at time $t = n$. 
The above model is the so-called \textit{basic collision model} \cite{Ciccarello2022QuantumInteractions}, and is a useful simple model amenable to rigorous analysis. 
In principal, however, the bath $\mcE$ may be subject to its own internal dynamics, which will effect both the unitary dynamics describing the joint system $\mcS\otimes\mcE$ and the state of $\mcE$ \cite{Attal2006FromInteractions, Grimmer2016OpenInteraction, Ciccarello2022QuantumInteractions}. 
To account for this, it is natural to instead consider the following modification of the above model: 
if the state of $\mcS$ is described by the density matrix $\arbrho_n$ at time $t = n$, then its state at time $t = n +1$ is given by 
\begin{equation*}
    \arbrho_{n+1}
    =
    \operatorname{Tr}_{\mcE}\big(U_n\seq{\arbrho_n\otimes\rho_n}U_{n}^*\big),
\end{equation*}
where it need not be the case that $U_n = U_{n-1}$ or $\rho_n = \rho_{n-1}$.
The present work allows us to consider the case that $\seq{U_n}_{n\in\mbN}$ and $\seq{\rho_n}_{n\in\mbN}$ are described by stationary ergodic stochastic processes, which recovers the following description of the \textit{random} open quantum dynamics of $\mcS$ so defined: 
if $\arbrho_0$ is the state of $\mcS$ at time $t = 0$, then $\arbrho_n$ is 
\begin{equation*}
    \arbrho_n
    =
    \phi_{n}\circ\cdots\circ\phi_1(\arbrho_0),
\end{equation*}
where $\seq{\phi_n}_{n\in\mbN}$ is a quantum channel-valued stochastic process satisfying an underlying ergodic regularity hypothesis.
This may either be interpreted as $\mcS$ interacting with various subsystems $\mcE_n$ of $\mcE$ subject to the internal, stochastic dynamics of $\mcE$, or as $\mcS$ interacting with a sequence of stochastic probes.
In this work, we seek to understand the basic dynamical properties of such random open quantum dynamical systems. 
\subsubsection{Quantum spin chains}
Quantum spin chains are one-dimensional models of many-body quantum systems that have been highly-studied over the past 30 years \cite{Fannes1992FinitelyChains}. 
In particular, the study of matrix product states on quantum spin chains has received a great deal of analysis. 
Recently, there has been an uptick in work analyzing states on disordered matrix product states, a research program spearheaded by the pair of papers \cite{Movassagh2021TheoryProcesses, Movassagh2022AnStates} and continued in the recent preprints \cite{Roon2025FinitelyDynamics, Pathirana2026CorrelationStates}.
Owed to the transfer operator description of such states described in \cite{Fannes1992FinitelyChains}, there is an ergodic quantum process $(\theta, E)$ with which one can analyze these disordered matrix product states, in which case solutions $\uniqrho\in\rstates$ to the equation $E_{\theta(\omega)}(\varrho_\omega) = \varrho_{\theta(\omega)}$ play a central role, as described in \cite{Movassagh2022AnStates}.
Heuristically, for a certain class of quantum states $\psi$ on the quantum spin chain, $(\theta, E)$ describes correlations between local observables $a, b$ on the quantum spin chain via 
\begin{equation*}
    \operatorname{Cor}_{\psi}(a, b)_\omega
    \sim 
    E^{(\operatorname{dist}(a, b))}_\omega,
\end{equation*}
where $\operatorname{dist}(a, b)$ denotes the distance between the supports of $a$ and $b$, and $\operatorname{Cor}_{\psi}(a, b)_\omega$ denotes the quantum mechanical correlation between $a$ and $b$ with respect to state $\psi$, given disorder configuration $\omega$. 
For more details, one may consult \cite{Roon2025FinitelyDynamics}.
\subsection{Background and relation to other work}
Ergodic quantum processes as we describe them here were introduced by Movassagh \& Schenker in \cite{Movassagh2021TheoryProcesses, Movassagh2022AnStates}, where they proved a strong ergodic theorem under an assumption of eventual strict positivity (ESP)---which, in the context of this work, may be understood as a particularly strong sort of irreducibility. 
In the deterministic setting $(\operatorname{Id}, \psi)$, this so-called ESP assumption is equivalent to the assumption that $\psi$ is irreducible and that the only eigenvalue of $\psi$ of modulus 1 is 1 itself. 
In a companion work \cite{Ekblad2026PeriodicityProcesses}, we study solutions $x\in\rmatrices$ to the eigenvalue equation equation 
\begin{equation*}
    \phi_{\theta(\omega)}(x_\omega)
    =
    \lambda x_{\theta(\omega)}
\end{equation*}
for $\lambda\in\mbC$ with $|\lambda| = 1$ under the assumption that $(\theta, \phi)$ is irreducible. 
Ergodic quantum processes were studied earlier under more specific stochastic assumptions, including i.i.d. and Markovian situations \cite{Bruneau2008RandomSystems, Bruzda2009RandomOperations, Bruneau2010InfiniteDynamics, Nechita2012RandomStates, Bougron2022MarkovianSystems}, or under specific algebraic assumptions, as in \cite{Ekblad2025AEntanglement}.
The general framework we put forth here unifies these previous works within a single theory while also serving as a vast generalization of the i.i.d. and Markovian frameworks, encompassing many other common models of randomness, including periodic, quasi-periodic, and some sequences of random variables with long-range correlations. 
We showcase this in more detail in our examples below (see \S \ref{Sec:Examples}).


Ergodic repeated interaction schemes may be viewed as encoding the expected behavior of an underlying quantum trajectory under random generalized measurements, which were studied in \cite{Ekblad2025AsymptoticMeasurements}. 
A central limit theorem for ESP ergodic quantum processes has been obtained in \cite{Pathirana2023LawProcesses}, and a large deviation principle is obtained in \cite{Schenker2024QuenchedMeasurements}.
In \cite{Ekblad2025AEntanglement}, a multiplicative ergodic theorem is proved for general ergodic quantum processes satisfying $\phi_\omega(\mbI) = \mbI$ almost surely. 
In \cite{Ekblad2026ErgodicMeasurements}, quantum trajectories associated to general, not necessarily ESP, dynamically ergodic ERISs are analyzed, and it is shown that the natural quantum measure associated to such trajectories is an invariant, ergodic measure for the shift map on the outcome space. 
Eventually strictly positive ergodic quantum processes on finite von Neumann algebras were studied in \cite{Nelson2024ErgodicAlgebras} and some ergodic results were obtained. 


What we call quantum processes here are also known as repeated interactions and/or quantum collision models in the literature. 
For a non-exhaustive list of relevant physics literature, one may consult \cite{Ciccarello2022QuantumInteractions, Cusumano2022QuantumGuide} and the references therein. 
On the mathematical physics side, one may consult  \cite{Attal2006FromInteractions, Bruneau2008RandomSystems, Pellegrini2009Non-MarkovianMeasurements, Burgarth2013ErgodicDimensions, Bruneau2014RepeatedSystems, Hanson2017LandauersSystems, Bougron2020LinearSystems, Zhang2024CriteriaDynamics} and the references therein.


The theory we present here may be understood as a natural extension of the Perron-Frobenius theory of positive operators on finite-dimensional $C^*$-algebras to the general statistically homogeneous setting, which, in turn, is itself an extension of the Perron-Frobenius theory of matrices with positive entries. 
The Perron-Frobenius theory of positive operators on Banach lattices and $C^*$-algebras has an extensive literature. 
The classic theorems of Perron and Frobenius may be found in \cite{Perron1907ZurMatrices, Frobenius1912UberElementen}, and are fully presented in \cite[Chapter 1]{Schaefer1974BanachOperators}. 
Perhaps the most well-known generalization of the Perron-Frobenius theory to the full generality of Banach spaces is the Krein-Rutman theorem \cite{Krein1948LinearSpace}. 
In general, the theory of positive operators on Banach lattices and positive operators on $C^*$-algebras is markedly different, as indicated by \cite{Kadison1951OrderOperators, Sherman1951OrderAlgebras}. 
Nevertheless, the general methodology used to understand the Perron-Frobenius theory of these two types of positive operators is similar.
The reader may find an excellent exposition of the Perron-Frobenius theory of positive operators on Banach lattices in \cite{Schaefer1974BanachOperators}, and other general discussions of positive operators on topological vector spaces in the latter half of \cite{Schaefer1999TopologicalSpaces}. 
For a list of literature related to the Perron-Frobenius theory of positive operators on $C^*$-algebras, see \cite{Stormer1963PositiveAlgebras, Choi1974AC-Algebras, Evans1978SpectralC-Algebras, Albeverio1978FrobeniusAlgebras, Enomoto1979AC-algebras, Groh1983OnC-Algebras, Groh1984UniformW-Algebras, Groh1984UniformlyC-algebras, Farenick1996IrreducibleAlgebras, Schrader2001Perron-FrobeniusIdeals} for a start. 


Our work may also be viewed as a specification of the Perron-Frobenius theory of positive linear cocycles on ordered Banach spaces $\scrX$ to the case where $\scrX$ is a finite-dimensional $C^*$-algebra with the canonical ordering. 
To our knowledge, the Perron-Frobenius theory of such cocycles has mostly been restricted to the specific case where $\scrX$ is a commutative finite-dimensional $C^*$-algebra satisfying some strong positivity assumptions, and the general noncommutative finite-dimensional $C^*$-algebra case without strict positivity has not received much attention outside of the above referenced literature.
For some results related to the above, the interested reader may consult \cite{Evstigneev1974PositiveSystems, Arnold1994EvolutionaryMatrices, Hennion1997LimitMatrices, Pollicott2010MaximalProducts, Mierczynski2013PrincipalTheory, Mierczynski2013PrincipalSystems}.
Specializing to the case of interest in quantum information specifically, the study of steady states, asymptotics, and decomposition theories has received attention for various reasons; the interested reader may consult the list \cite{Arias2002FixedOperations, Li2011FixedOperations, Li2011CharacterizationsOperations, Burgarth2013ErgodicDimensions, Carbone2016OpenProperties, Carbone2020OnChannel, Carbone2021AbsorptionChannels} for some further reading on this topic.
%


For an exposition of the theory of measure-preserving ergodic transformations, see \cite{Walters1982AnTheory}. 
There, one may also find an exposition of the spectral theory of ergodic transformations and the relationship between this and the mixing properties of the transformation. 
For a more operator-theoretic perspective on ergodic theory, one may consult \cite{Eisner2015OperatorTheory}. 
%
%
%
%


%
%
%
%
\subsection{Organization}
In \S \ref{Sec:Examples}, we give some relevant examples to illuminate our theory. 
We focus in particular on how our results relate to existing literature on random repeated interactions, in particular, the i.i.d. and Markovian models. 
In \S \ref{Sec:Reducibility theory}, we give all the main technical results needed to prove Theorems \ref{Thm:Minimal projections}, \ref{Thm:Dynamically ergodic theorem}, and \ref{Thm:Recurrence}, and then we give proofs of these theorems.
Lastly, in \S \ref{Sec: IID theorem}, we prove Theorem \ref{Thm:I.I.D. processes}. 
%
%
%
%


\section{Examples}\label{Sec:Examples}
We now illustrate the above theory with some examples. 
\subsection{Independent and identically distributed}
As advertised above, our theory gives a tool to study products of independent and identically distributed (i.i.d.) quantum channels, as we now describe. 
Let $(\Xi, \mcG, \nu)$ be a probability space, and let $\psi:\Xi\to\channels$ be a random quantum channel.
We let $\Omega$ be the probability space 
\[
    \Omega = \prod_{i\in\mbZ}\Xi
\]
with the product $\sigma$-algebra and the product measure $\mu = \bigotimes_{i\in\mbZ}\nu$.
Let $\theta:\Omega\to\Omega$ be the left shift map, i.e., $\theta: (\omega_i)_{i\in\mbZ}\mapsto (\omega_{i+1})_{i\in\mbZ}$. 
It is a standard fact that $\theta$ is an invertible ergodic measure preserving transformation, so, in particular, $(\theta, \phi)$ defines an ergodic quantum process, 
and the sequence of quantum channels $\seq{\phi_n}_{n\in\mbZ}$ given by 
\begin{align*}
    \phi_n:\Omega&\to\channels\\
    (\omega_i)_{i\in\mbZ}&\mapsto \phi_n =  \psi_{\omega_n} = \psi_{\theta^n(\omega_0)}
\end{align*}
is an independent, identically distributed (\textit{i.i.d.}) sequence of random quantum channels. 
We call an ergodic quantum process obtained in this way an \textit{i.i.d. quantum process}, and, for such processes, we have the following general fact.
We say $x\in\rmatrices$ is deterministic if there is a matrix $m\in\matrices$ such that $x_\omega = m$ for almost every $\omega\in\Omega$. 
\begin{restatable}[]{thm}{IIDTheorem}
    \label{Thm:I.I.D. processes}
Let $(\theta, \phi)$ be an i.i.d. ergodic quantum process. 
If the recurrent projection $\recP$ is deterministic, then any minimal reducing projection for $\seq{\phi_n}_{n\in\mbZ}$ is deterministic.
\end{restatable}
The recurrent projection is deterministic, for example, if $\phi(\mbI) = \mbI$ almost surely, or if $(\theta, \phi)$ is irreducible. 
We now consider some specific examples of i.i.d. quantum processes. 
\begin{example}[Haar unitary processes]
    Using the above notation, let $\seq{\Xi, \mcG, \nu} = \seq{\mbU_d, \mcB(\mbU_d), \operatorname{Haar}}$, where $\mbU_d$ is the set of $d\times d$ unitary matrices, $\mcB(\mbU_d)$ is the Borel $\sigma$-algebra on $\mbU_d$, and $\operatorname{Haar}$ is the Haar measure on $\mbU_d$. 
    Then let $\psi:\mbU_d\to\bops{\matrices}$ be given by 
    \begin{equation}
        \psi_u(\cdot) := u(\cdot)u^*.
    \end{equation}
    We call the resulting i.i.d. quantum process $(\theta, \psi)$ the ($d$-dimensional) Haar unitary process, and we denote it by $\mfU$ for shorthand. 
    We claim that $\mfU$ is irreducible. 
    To see this, let $p$ be a nontrivial minimal reducing projection for $\mfU$. 
    It is clear that $\phi_n(\mbI) = \mbI$ for all $n$, so by Theorem \ref{Thm:I.I.D. processes}, $p$ is deterministic. 
    Thus, to show that $\mfU$ is irreducible, by Lemmas \ref{Lem:Equivalent characterizations of reducibility} and \ref{Lem:Min red proj fixed for recurrent corner} below, it suffices to show that if $p\neq 0$, then
    \begin{equation}
        upu^* = p
    \end{equation}
    for $\operatorname{Haar}$-almost every $u\in\mbU_d$ implies $p = \mbI$.
    To see this, consider the map
    \begin{equation*}
        \begin{split}
            F: \mbC^d\times\mbC^d&\to\mbC\\
        \big(\ket{\varphi}, \ket{\varphi'}\big)
        &\mapsto 
        \bra{\varphi}p\ket{\varphi'}.        
        \end{split}
    \end{equation*}
    Then by assumption on $p$, we have that 
    \begin{equation}
        F\big(\ket{\varphi}, \ket{\varphi'}\big)
        =
        \int_{\mbU_d}
        \bra{\varphi}upu^*\ket{\varphi'}\,
        d\nu(u).
    \end{equation}
    So, $F$ defines a positive-semidefinite Hermitian form invariant under the unitary action of $\mbU_d$ on $\mbC^{d}\otimes \mbC^{d}$ defined by $w\big(\ket{\varphi}, \ket{\varphi'}\big) = \big(w\ket{\varphi}, w\ket{\varphi'}\big)$. 
    In particular, the value of $F(\ket{\varphi},\ket{\varphi})$ is independent of $\ket{\varphi}$, and is consequently equal to either 0 or 1 for all $\ket{\varphi}\in\mbC^d$, which implies that $p = \mbI$ or $p = 0$. 
    This is what we wanted to show, hence $\mfU$ is irreducible. 
    As a consequence of this irreducibility, we may easily deduce some facts about $\mfU$ that have more or less already appeared in the literature (see, e.g., \cite{Nechita2012RandomStates}). 
    For example, we see that the unique invariant state for $\mfU$ is $\mbI$, so by the irreducibility of $\mfU$, we conclude from Theorem \ref{Thm:Dynamically ergodic theorem} that 
    \begin{equation*}
        \lim_N\frac{1}{N}
        \sum_{n=1}^N
        u_n\cdots u_1 \arbrho u_1^*\cdots u_n^*
        =
        d^{-1}\mbI 
    \end{equation*}
    for any measurable $\arbrho:\Omega\to\states$. 
    In particular, $\arbrho$ may be arbitrarily correlated with $(u_n)_{n\in\mbN}$.
\end{example}
\begin{example}[Random repeated interaction quantum systems \cite{Nechita2012RandomStates, Bruneau2008RandomSystems}]\label{Ex:Repeated quantum interactions}
    Let $(\theta, \phi)$ be an i.i.d. quantum process. 
    Assume as in \cite[Theorem 5.2]{Nechita2012RandomStates} that
    \begin{equation*}\label{Eqn:Ex:Repeated quantum interactions, Eqn 1}
        \nu\seq{\text{there is a unique $\rho\in\states$ such that for $\phi(\rho) = \rho$}} > 0.
    \end{equation*}
    Let $\mfI$ denote the corresponding i.i.d. quantum process. 
    If we assume further that $\recP$ is deterministic, then we have that $\recP$ is the unique minimal reducing projection for $\mfI$, i.e., $\mfI$ is dynamically ergodic. 
    Indeed, if not, then by Theorem \ref{Thm:I.I.D. processes}, there are two nonzero deterministic projections $p_1, p_2\in\matrices$ such that $p_k$ reduces $\mfI$ for both $k$ and $p_1\neq p_2$, which we assume without loss of generality are both minimal.
    So, $\phi_x(p_k\matrices p_k)\subseteq p_k\matrices p_k$ for almost every $x\in\Xi$, so that $p_k$ reduces $\phi$ almost surely.
    In particular, for both $k$, almost surely there are states $\rho_k\in\states$ such that $\operatorname{ran}(\rho_k) = \operatorname{ran}(p_k)$ and $\phi(\rho_k) = \rho_k$ via \cite{Evans1978SpectralC-Algebras}. 
    Thus, 
    \begin{equation*}
          \nu\seq{\text{there is a unique $\rho\in\states$ such that  $\phi(\rho) = \rho$}} = 0,
    \end{equation*}
    a contradiction to our assumption.
    So, $\recP$ is a minimal reducing projection. 
    Therefore, Theorems \ref{Thm:Minimal projections} and \ref{Thm:Dynamically ergodic theorem} imply there is unique $\uniqrho\in\rstates$ such that for all $\arbrho\in\rstates$, we have the $\mu$-almost sure convergence 
    \begin{equation}\label{Eqn:Example:mbE varrho}
        \lim_{N\to\infty}
        \frac{1}{N}
        \sum_{n=1}^{N}
        \phi_n\circ\cdots\circ\phi_1(\arbrho)
        =
       \mbE_\mu[\uniqrho]
    \end{equation}
    which offers an extension of the convergence result \cite[Theorem 5.2]{Nechita2012RandomStates}, in the present situation where $\recP$ is deterministic. 
    In particular, the convergence in \cite[Theorem 5.2]{Nechita2012RandomStates} holds for not just deterministic input states $\arbrho$, but any random input states $\arbrho\in\rstates$, which may be arbitrarily correlated with $\seq{\phi_n}_{n\in\mbN}$.
    Moreover, from \cite[Theorem 5.2]{Nechita2012RandomStates}, we know that the fixed point space of the quantum channel $\mbE[\phi]$ is $1$-dimensional, i.e., there is unique $\rho\in\states$ such that $\mbE_\mu[\phi](\rho) = \rho$, so by taking expectation of (\ref{Eqn:Example:mbE varrho}), we conclude that $\rho = \mbE_\mu[\uniqrho]$.
\end{example}
\subsection{Other examples}
\begin{example}[Markovian repeated interaction quantum systems \cite{Bougron2022MarkovianSystems}]\label{Example:Markovian}
In \cite{Bougron2022MarkovianSystems}, the authors considered \textit{Markovian repeated interaction quantum systems (MRIS)}, and the authors of \cite{Bougron2022MarkovianSystems} proved ergodic theorems \cite[Theorem 3.2 \& Theorem 3.3]{Bougron2022MarkovianSystems}. 
We now compare their set-up with our framework, so that in particular the reader may compare their results \cite[Theorem 3.2 \& Theorem 3.3]{Bougron2022MarkovianSystems} with our Theorems \ref{Thm:Minimal projections} and \ref{Thm:Dynamically ergodic theorem}.
Let $\Xi$ be a finite set, and enumerate the points of $\Xi$ as $\Xi = \set{\xi_0, \dots, \xi_{m-1}}$.
Let $\Gamma$ be a directed graph whose vertices $V_\Gamma$ are the points of $\Xi$ and whose edges $E_\Gamma$ make $\Gamma$ into a strongly connected graph. 
For each $(\xi_i, \xi_j)\in E_\Gamma$, let $p_{(j, i)}\in (0, 1)$ be a number such that for all $\xi_i$ we have that 
\begin{equation}
    \sum_{\xi_j\in\Xi\,\,:\,\, (\xi_i, \xi_j)\in E_\Gamma}
    p_{(j, i)}
    =
    1.
\end{equation}
This defines an irreducible Markov chain $X$ with state space $\Xi$ by the rule 
\begin{equation}
    \operatorname{Prob}(X = \xi_j\,\,| \,\, X = \xi_i)
    =
    p_{(j, i)},
\end{equation}
so, in particular, if we let $\boldsymbol{p}$ be the unique distribution on $\Xi$ that is invariant under the action induced by this Markov chain, then it is a standard fact that this induces a probability measure $\mbP$ on the space $\Xi^\mbZ$ that is ergodic with respect to the map shift $T:\Xi^\mbZ\to\Xi^\mbZ$ defined by $T\big(\seq{\xi_{i_k}}_{k\in\mbZ}\big) = \seq{\xi_{i_{k+1}}}_{k\in\mbZ}$.
So, if we let $\set{\psi_{i}}_{i=0}^{m-1}$ be a collection of quantum channels, then 
\begin{equation}
    \begin{split}
        \phi:\Xi^\mbZ &\to \{\text{quantum channels}\}\\
        \seq{\xi_{i_k}}_{k\in\mbZ}&\mapsto \psi_{i_0}
    \end{split}
\end{equation}
defines an ergodic quantum process $(T, \phi)$.
Note that our ergodicity assumption amounts to assuming the underlying Markovian process is \textit{irreducible}.
This assumption is sufficient to conclude the results \cite[Theorem 3.2 \& Theorem 3.3]{Bougron2022MarkovianSystems}, but \cite{Bougron2022MarkovianSystems} prove their theorems for a larger class of underlying Markov processes. 
\end{example}
\begin{example}[Reducible ergodic quantum processes from deterministic irreducible channels]
We now construct an example of a reducible ergodic quantum process where we can explicitly describe the minimal reducing projections of Theorem \ref{Thm:Recurrence}, and we give the stationary states corresponding to these minimal reducing projections of Theorem \ref{Thm:Minimal projections}.
To do this, let $\psi\in\channels$ be a fixed irreducible quantum channel, and suppose that $|\operatorname{PerSpec}(\psi)| = h > 1$, where $\operatorname{PerSpec}(\psi)$ is the set of eigenvalues $\lambda\in\mbC$ of $\psi$ with $|\lambda| = 1$.
By \cite{Evans1978SpectralC-Algebras}, we know that $\operatorname{PerSpec}(\psi)$ is equal to $\Gamma_h$, the set of $h$th roots of unity. 

Now set $\Omega = \Gamma_h$ and let $\mu$ be the uniform probability distribution on $\Omega$, where $\Omega$ has the discrete $\sigma$-algebra. 
Let $\theta:\Omega\to\Omega$ be the map $\omega\mapsto \alpha \omega $, with $\alpha = e^{2\pi i/h}$.
It is then clear that $\theta$ is an invertible ergodic measure preserving transformation, so if we let $\phi:\Omega\to\channels$ be $\phi_\omega = \psi$, then $(\theta, \phi)$ defines an ergodic quantum process; for simplicity, we denote this by $(\theta, \psi)$.

We now show that $\recP = \mbI$ and that $(\theta, \psi)$ is a reducible ergodic quantum process, and we explicitly construct the partition of identity that exists as guaranteed by Theorem \ref{Thm:Minimal projections}.
In particular, we show that this gives an example of a deterministic quantum channel $\psi\in\channels$ where $(\operatorname{Id}, \psi)$ is an irreducible ergodic quantum process, but there exists a probability space $(\Omega, \mu)$ and a measure preserving transformation $\theta:\Omega\to\Omega$ such that $(\theta, \psi)$ is a reducible ergodic quantum process. 
First, we show that $(\theta, \psi)$ is reducible. 
To see this, let $Z_\alpha\in\matrices$ with $\psi(Z_\alpha) = \alpha Z_\alpha$.
So, if we let $f_\alpha:\Omega\to\mbC$ be the identity function $f_\alpha(\omega)=\omega$, then
we see that $f_\alpha(\theta^{-1}(\omega)) = \alpha^{-1}\omega = \alpha^{-1}f_\alpha(\omega)$ for all $\omega\in\Omega$. 
Thus, $X:= f_\alpha Z_\alpha\in\rmatrices$ satisfies 
\begin{equation*}
    \phi_\omega\!\seq{X_{\theta^{-1}(\omega)}} 
        =
    f_\alpha(\theta^{-1}(\omega))
    \psi(Z_\alpha)
        =
    X_\omega,
\end{equation*}
so that $X\in\rmatrices$ is contained in the vector space $\operatorname{Fix}_\mbI$ defined in Theorem \ref{Thm:Minimal projections}. 
However, clearly $X\not\in\rstates$, so $\operatorname{Fix}_\mbI$ is at least two-dimensional. Hence by Theorem \ref{Thm:Minimal projections}, it holds that $(\theta, \psi)$ is reducible. 
Now, since $(\theta, \psi)$ is reducible, there must exist $p\in\scrP_{(\theta, \psi)}$ with $p\neq\mbI$. 
We can explicitly construct such a $p$: 
from \cite[Theorem 4.2]{Evans1978SpectralC-Algebras}, there are (nonzero, nonidentity) deterministic projections $p_k$ for each $k\in\mbZ/h\mbZ$ such that $\psi(p_{k}\matrices p_k)\subseteq p_{k+1}\matrices p_{k+1}$ for all $k$. 
Thus, if we define 
\begin{equation*}
    p_\omega 
    := 
    \sum_{k\in\mbZ/h\mbZ}
    1_{\set{e^{2\pi i k/h}}}(\omega)p_k,
\end{equation*}
then we see from construction that $\psi(p_\omega\matrices p_\omega)\subseteq p_{\theta(\omega)}\matrices p_{\theta(\omega)}$ holds for all $\omega\in\Omega$. 
So, clearly $p$ reduces $(\theta, \psi)$ and $p\neq \mbI$. 

We now show that $p$ is minimal. 
To see this, we apply Theorem \ref{Thm:Minimal projections} (d).
Let $\arbrho\in\rstates$, let $x\in\mbP_d^\Omega$ be such that $x_{\omega_0}\neq 0$ for some $\omega_0$, and suppose that $p\arbrho p = \arbrho$ and $pxp = x$. 
Suppose that $\omega_0 = e^{2\pi i k/h}$, and note that $p_{\omega_0}  =p_k$. 
Notice that $\phi^{(h)}_\omega = \psi^h$ satisfies 
$$\phi^{(h)}_\omega(p_k\matrices p_k)\subseteq p_{\theta^h(\omega)}\matrices p_{\theta^h(\omega)} = p_k\matrices p_k \ .$$
Because $p_k$ is a minimal reducing projection of $\psi^h$ and $\psi^h\vert_{p_k\matrices p_k}$ satisfies $\operatorname{PerSpec}(\psi^h\vert_{p_k\matrices p_k}) = \set{1}$ via the proof of \cite[Theorem 4.4]{Evans1978SpectralC-Algebras}, we conclude from \cite[Ch. 6]{Wolf2012QuantumTour} that there is $\rho_k\in p_k\mbP_d p_k$ with $\operatorname{Tr}(\rho_k) = 1$ and $\operatorname{ran}(\rho_k) = \operatorname{ran}(p_k)$ such that $\lim_{n} \psi^{hn}(a)=\operatorname{Tr}(a)\rho_k$ for all $a\in p_k\matrices p_k$. Thus, for $\omega=e^{2\pi i j/h}$, $\lim_{n}\phi_\omega^{(k-j)+nh}(\eta_\omega) = \rho_k$, so there is $N_j\ge 0$ such that for $n\ge N_j$
\begin{equation*}
    \operatorname{Tr}(
        \phi^{(k-j)+nh}_\omega(\arbrho_\omega)
        x_{\theta^{(k-j)+nh}(\omega)}     )
 \ = \  \operatorname{Tr}(
        \phi^{(k-j)+nh}_\omega(\arbrho_{\omega}) 
        x_{\omega_0}
    )
\    >  \ 0 \ .
\end{equation*}
%
As this holds for each $\omega\in \Omega$, it follows from condition (d) of Theorem \ref{Thm:Minimal projections} that $p$ is indeed minimal. 
By a similar argument, we see that $p^{(j)}\in\rmatrices$ defined by $p^{(j)}_\omega := p_{\theta^j(\omega)}$ is a minimal reducing projection for $(\theta, \psi)$ for all $j = 0, \dots, h - 1$, and clearly $\sum_j p^{(j)} = \mbI$ holds. 
Therefore, $\set{p^{(j)}}_{j=0}^{h-1}$ is a set of minimal reducing projections as in Theorem \ref{Thm:Recurrence}. 
It is clear from inspection that the states $\varrho^{(j)}\in\rstates$ corresponding to $p^{(j)}$ as in Theorem \ref{Thm:Minimal projections} are $\varrho^{(j)}_{\omega} 
    := 
    \rho_{k + j}$
when $\omega = e^{2\pi i k/h}$, which holds since $\psi(\rho_k) = \rho_{k+1}$ for all $k\in\mbZ/h\mbZ$ via \cite[Theorem 4.4]{Evans1978SpectralC-Algebras}.
This is all we wanted to show. 
\end{example}
It is not hard to see from this discussion that the following proposition holds. 
Recall that $\psi\in\channels$ is called \textit{primitive} if $\operatorname{PerSpec}(\psi) = \{1\}$. 
\begin{prop}
    Let $\psi\in\channels$. 
    Then $\psi$ is primitive if and only if $(\theta, \psi)$ is an irreducible ergodic quantum process for all probability spaces $(\Omega, \mu)$ and invertible ergodic measure preserving transformations $\theta:\Omega\to\Omega$. 
\end{prop}
\begin{proof}
    If $\psi$ is not primitive, then the above example shows that there exists $(\Omega, \mu)$ and $\theta:\Omega\to\Omega$ such that $(\theta, \psi)$ is reducible. 
    Conversely, if $\psi$ is primitive, then there is a full-rank state $\rho\in\states$ uch that $\lim_n \psi^n(x) = \operatorname{Tr}(x)\rho$ for all $x\in\matrices$ \cite[Ch. 6]{Wolf2012QuantumTour}. 
    Therefore, for any probability space $(\Omega, \mu)$ and invertible ergodic measure preserving transformation $\theta:\Omega\to\Omega$, if $p_\omega$  is a non-trivial random projection, then $\psi^n(p_\omega)$ is eventually full rank with probability one.  It follows that the only non-trivial reducing projection is $p_\omega = \mbI$, 
    so $(\theta, \psi)$ is irreducible, concluding the proof. 
\end{proof}
%
%
%
%


\section{Reducibility theory}\label{Sec:Reducibility theory}
In this section, we provide the bulk of the technical reducibility theory. 
First, we set notation and terminology.
\subsection{Notation, terminology, and conventions}\label{Subsec:Notation}
We begin by reiterating some of the notation that was set in the introduction in addition to setting new notation.
We let $\matrices$ denote the set of $d\times d$ matrices with entries in $\mbC$. 
For $a\in\matrices$, we let $\sigma(a)$ denote the set of eigenvalues of $a$. 
For $a\in\matrices$, we let $a^*$ denote the conjugate transpose of $a$. 
We let $\mbP_d\subset\matrices$ denote the set of positive semidefinite matrices, and to indicate that $a\in\mbP_d$, we write $a\geq 0$ or $0\leq a$, and we say $a$ is positive (dropping the word semidefinite).
For any $a\in\matrices$, we let $|a|\in\mbP_d$ be $\sqrt{a^*a}$ as defined by the functional calculus. 
We say $p\in\matrices$ is a projection if $p = p^* = p^2$. 
For any $a\in\matrices$, we write $\proj{a}$ to denote the projection onto $\operatorname{ran}(a)$. 
We write $\mbI$ to denote the identity in $\matrices$. 
We let $\tr{a}$ denote the usual trace of the matrix $a\in\matrices$. 
We write $\states$ to denote the set of $\rho\in\mbP_d$ with $\tr{\rho} = 1$ and we call $\rho\in\states$ a state.
We say $a\in\matrices$ is self-adjoint if $a^* = a$. 
We make $\matrices$ into a Hilbert space by means of the Hilbert-Schmidt inner product, which we recall is the inner product defined by 
\begin{equation*}
    \inner{a}{b}
    =
    \tr{a^*b}\text{ for all $a, b\in\matrices$.}
\end{equation*}
Note that we have taken the convention that $\inner{\cdot}{\cdot}$ is conjugate linear in the first argument and linear in the second. 
For any $p\in [1, \infty)$, we let $\|\cdot\|_p$ be the norm on $\matrices$ defined by 
\begin{equation*}
    \|a\|_p
    := 
    \tr{|a|^p}^{1/p}.
\end{equation*}
In particular, $\|a\|_2 = \inner{a}{a}^{1/2}$.
We define $\|\cdot\|_\infty$ to be the operator norm of $A$ viewed as a bounded linear operator on $\mbC^d$ when $\mbC^d$ is made into a Hilbert space in the usual way. 
For a Banach space $\mcX$, let $\bops{\mcX}$ denote the space of bounded operators on $\mcX$, and we let $\dual{\mcX}$ denote the dual of $\mcX$, i.e., the set of bounded linear functionals $f:\mcX\to\mbC$.\footnote{We have opted to use the symbol $\star$ to denote the dual so as to more easily discern it from the adjoint $*$.}
When we refer to a measurability structure on a finite-dimensional Banach space, we are always referring to the Borel $\sigma$-algebra structure on it defined by any complete norm. 
We call an element of $\superops$ a \textit{superoperator}. 
We say a superoperator $\psi\in\superops$ is \textit{positive} if $\psi(\mbP_d)\subseteq\mbP_d$, and we write $\positivesuperops$ to denote the set of positive elements of $\superops$. 
%
%
For any $\psi\in\superops$, let $\psi^*\in\superops$ be the Hilbert space adjoint, i.e., the unique element of $\superops$ satisfying
\begin{equation*}
    \inner{\phi(a)}{b}
    =
    \inner{a}{\phi^*(b)}
\end{equation*}
for all $a, b\in\matrices$. 
Note that the adjoint $\phi^*$ of a positive map $\phi\in\superops$ is itself positive. 
We say that $\psi\in\superops$ is completely positive if for all $n\in\mbN$, the map $\psi\otimes\operatorname{Id}_n\in\bops{\matrices\otimes\mbM_n}$ is positive, i.e., for all $a\in\matrices\otimes\mbM_n$, $\psi\otimes\operatorname{Id}_n(a^*a)\geq 0$, where $\operatorname{Id}_n\in\bops{\mbM_n}$ is the identity. 
It is clear that the adjoint of a completely positive map is, itself, completely positive. 
We say that $\psi\in\superops$ is {trace-preserving} if for all $a\in\matrices$, $\tr{\psi(a)} = \tr{a}$. 
It is straightforward to show that this is equivalent to $\psi^*$ being unital, i.e., that $\psi^*(\mbI) = \mbI$. 
We denote the set of all completely positive and trace preserving maps $\psi\in\superops$ by $\channels$, and we call an element of $\channels$ a \textit{quantum channel}. 
The various norms on $\matrices$ induce various norms on $\superops$.
Given a norm $|||\cdot|||$ on $\matrices$ and a linear map $\psi:\matrices\to\matrices$, we write $|||\psi|||$ to denote the norm 
\begin{equation*}
    |||\psi|||
        := 
    \sup_{|||a||| = 1}|||\psi(a)|||.
\end{equation*}
In this work, we are concerned with random elements of $\matrices$ and $\superops$, i.e., measurable functions $\Omega\to\matrices$ and $\Omega\to\superops$, so let us set notation for such objects. 
For a probability space $(\Omega, \mcF, \mu)$ with $\sigma$-algebra $\mcF$ and measure $\mu$, we say a measurable map $T:\Omega\to\Omega$ satisfying $\mu[T^{-1}(F)] = \mu[F]$ for all $F\in\mcF$ is a measure-preserving transformation, which we abbreviate by \textit{mpt.}
We call $T$ invertible only if its inverse is also measurable, and, if $T$ is an mpt., we call $T$ ergodic if for all $F\in\mcF$, $T^{-1}(F) = F$ implies $\mu[F]\in\set{0, 1}$. 
If $(\Xi, \mcG)$ is a measure space and $f:\Omega\to\Xi$ is measurable, we write $\koopman{T}(f):\Omega\to\Xi$ to denote the measurable map defined by $\koopman{T}(f)(\omega) = f(T(\omega))$. 
It is clear that the restriction of $\koopman{T}$ to $L^2(\Omega)$ is unitary whenever $T$ is an invertible mpt.
We write $\mbE_\mu[\cdot]$ to denote $\int_\Omega(\cdot)\,\dee\mu$, sometimes writing $\mbE[\cdot]$ when $\mu$ is clear from context. 
We write $\rmatrices$ to denote the set of random matrices, i.e., the set of measurable functions $x:\Omega\to\matrices$, up to $\mu$-almost sure equivalence. 
Note that this set is a vector space over $\mbC$. 
We say that $x\in\rmatrices$ is a projection (resp. self-adjoint, unitary, positive, etc.) if $x$ is a projection almost surely (resp. self-adjoint, unitary, positive, etc., almost surely). 
We write $\rpositives$ to denote the set of positive elements $x\in\rmatrices$, and we write $\rstates$ to denote the set of states $\rho\in\rmatrices$. 
We say $x\in\rmatrices$ is nonzero if $\mu[x = 0] < 1$.
For $x\in\rmatrices$, if there is $a\in\matrices$ such that $x = a$ almost surely, we say $x$ is deterministic and simply write $x = a$.
For instance, we write $\mbI\in\rmatrices$ to denote the element of $\rmatrices$ that is equal to $\mbI$ almost surely, recalling that elements of $\rmatrices$ are only defined up to almost sure equivalence. 
Given a projection $p\in\rmatrices$, we write $\Prmatrices$ to denote the subspace of $\rmatrices$ given by 
\begin{equation*}
    \Prmatrices
    :=
    \set{x\in\rmatrices\,\,:\,\, pxp = x\text{ almost surely}}.
\end{equation*}
Given a linear map $\mfM:\rmatrices\to\rmatrices$, we say that $\mfM$ is positive if $\mfM(\rpositives)\subseteq\rpositives$, and we say $\mfM$ is completely positive if $\mfM\otimes\operatorname{Id}_n:\rmatrices\otimes\mbM_n\to\rmatrices\otimes\mbM_n$ is positive for all $n$, where we identify the algebraic tensor product $\rmatrices\otimes\mbM_n$ with $\mbM_{dn}^\Omega$ in the natural way.
We encounter a particular example of such maps in this work: 
given a random linear mapping $\phi:\Omega\to\superops$, we define the linear map $M_\phi:\rmatrices\to\rmatrices$ by 
\begin{equation*}
    M_\phi(x):\Omega\to\matrices,
        \quad  
    M_\phi(x)_\omega
        := 
    \phi_\omega(x_\omega).
\end{equation*}
It is clear that if $\phi$ is positive (resp. completely positive) almost surely, then $M_\phi$ is positive (resp. completely positive).
Given $p, q\in[1, \infty]$, we let $\Lpqmatrices$ denote the subset of $\rmatrices$ defined by 
\begin{equation*}
    \Lpqmatrices
    :=
    \set{
        x\in\rmatrices\,\,:\,\,
    \|x\|_q\in L^p(\Omega)
    },
\end{equation*}
which is a Banach space when given the norm $\|x\|_{\Lpqmatrices}:=\|\,  \|x\|_q\,  \|_{L^p(\Omega)}$. 
In the case that $p = q$, we simply write $\Lpmatrices$ to denote $L^p_p(\Omega, \matrices)$. 
It is useful to note that when $p\in [1, \infty)$, it holds that $\dual{\Lpqmatrices} = L^{p'}_{q'}(\Omega, \matrices)$, where $p'$ (resp $q'$) is the H\"older conjugate of $p$ (resp. $q$).
In particular, $\Ltwomatrices$ is a Hilbert space with inner product $\inner{\cdot}{\cdot}_{\Ltwomatrices} := \mbE_\mu\inner{\cdot}{\cdot}$.
%
%
%
%
%
\subsection{Reducing projections and minimality}\label{Subsec:Red proj and minimality}
For the rest of this work, $\seq{\Omega, \mcF, \mu}$ will be a fixed probability space, where $\mcF$ denotes the $\sigma$-algebra and $\mu$ denotes the measure. 
Given a measurable map $T:\Omega\to\Omega$ and a random superoperator $\psi:\Omega\to\superops$, we call the tuple $(T, \psi)$ a superoperator process. 
    Given a superoperator process $(T, \psi)$ and $n\in\mbN$, we let $\psi^{(n)}_{T}:\Omega\to\superops$ denote the random superoperators defined by 
    \begin{equation*}
        \psi^{(n)}_{T; \omega}:= \psi_{T^n(\omega)}\circ\cdots\circ\psi_{T(\omega)}.
    \end{equation*}
    If $T$ is clear from context, we write $\psi^{(n)}$ to denote $\psi^{(n)}_T$. 
    For a superoperator process $(T, \psi)$, we define the linear map $\glop{T}{\psi}:\rmatrices\to\rmatrices$ by
    \begin{equation*}
    \glop{T}{\psi}:\rmatrices\to\rmatrices,\quad 
    \glop{T}{\psi}(x)_\omega   
        := 
    \psi_{T(\omega)}(x_{T(\omega)}).
    \end{equation*}
    Notice that $\glop{T}{\psi}^n(x)_\omega = \psi^{(n)}_{\omega}(x_{T^n(\omega)})$.
If $T$ is invertible, we write $(T, \psi)^\dagger$ to denote the superoperator process $(T^{-1}, \koopman{T}(\psi))$, and we let $\glopdag{T}{\psi}:\rmatrices\to\rmatrices$ denote the linear map $\glop{T^{-1}}{\koopman{T}(\psi)}$, i.e., 
\begin{equation*}
    \glopdag{T}{\psi}(x)_\omega 
    =
    \psi_{\omega}(x_{T^{-1}(\omega)}).
\end{equation*}
Notice that if $T$ is an invertible mpt., then
\begin{equation*}
    \int_\Omega \inner{\glop{T}{\psi}(x)}{y}\,\dee\mu 
        =
    \int_\Omega \inner{x}{\glopdag{T}{\psi}(y)}\,\dee\mu 
\end{equation*}
for all $x, y\in\rmatrices$ for which the above integrals are absolutely convergent. 
In particular, if $\glop{T}{\psi}\vert_{\Ltwomatrices}\in\bops{\Ltwomatrices}$, then the adjoint of this bounded operator is $\glopdag{T}{\psi}\vert_{\Ltwomatrices}$.
We use this fact freely in the following. 
Given a superoperator process $(T, \psi)$, we say $(T, \psi)$ is positive (resp. trace preserving or quantum) if $\psi\in\positivesuperops$ (resp. $\psi$ is trace preserving or $\psi\in\channels$) almost surely. 
We say $(T, \psi)$ is stationary (resp. ergodic) if $T:\Omega\to\Omega$ is an invertible mpt. (resp. invertible ergodic mpt.).
Given that $(T, \psi)$ is positive (resp. stationary, trace preserving, quantum, or ergodic) we simply call $(T, \psi)$ a positive process (resp. stationary, trace preserving, quantum, or ergodic process), dropping the word superoperator. 
In particular, if, e.g., $(T, \psi)$ is ergodic and quantum, we call $(T, \psi)$ an ergodic quantum process.\footnote{For another example, if $(T, \psi)$ is stationary, positive, and trace preserving, we call $(T, \psi)$ a stationary positive trace preserving process.}
We reserve the symbol $\theta:\Omega\to\Omega$ for invertible ergodic mpt. and the symbol $\phi$ for positive trace preserving maps, so that the notation $(\theta, \phi)$ is reserved for ergodic positive trace preserving processes.
\begin{definition}[Reducing projections]\label{Def:Reducing projections}
    We say a random projection $p\in\rmatrices$ reduces a positive linear map $\mfM:\rmatrices\to\rmatrices$ if $\mfM(\prmatrices)\subseteq\prmatrices$, in which case we say that $p$ is a reducing projection for $\mfM$. 
    We say a reducing projection $p$ is nontrivial if $\mu[p = 0] < 1$. 
    We let $\RedProj{\mfM}$ denote the set of nontrivial reducing projections of $\mfM$. 
    If $\RedProj{\mfM}= \set{\mbI}$, we call $\mfM$ irreducible. 
    In the case that $\mfM = \glop{T}{\psi}$ for a positive process $(T, \psi)$, we sometimes say that $p$ reduces $(T, \psi)$ instead of $\glop{T}{\psi}$, and we write $\RedProj{(T, \psi)}$ to denote the set of nontrivial reducing projections of $(T, \psi)$. 

\end{definition}
\begin{remark}\label{Rmk:Relationship between operator and process}
    In the introduction, we said that a projection $p\in\rmatrices$ reduced an ergodic quantum process $(\theta, \phi)$ if $\phi_{\theta(\omega)}(p_\omega\matrices p_\omega)\subseteq p_{\theta(\omega)}\matrices p_{\theta(\omega)}$ held almost surely, which is slightly different from the above definition. 
    These two notions are essentially equivalent (up to duality, see Lemma \ref{Lem:L is reduced by P iff L dagger is reduced by I - P} below), however, as one can check that $p$ reduces $(\theta, \phi)$ according to the definition in the introduction if and only if $p$ reduces $(\theta^{-1}, \koopman{\theta}(\phi))$ according to Definition \ref{Def:Reducing projections}.
    The reason we use Definition \ref{Def:Reducing projections} as our main definition is because it is better suited to deal with the case that $T$ is not invertible. 
\end{remark}
The following lemma is standard.
\begin{lemma}\label{Lem:Equivalent characterizations of reducibility}
    Let $(T, \psi)$ be a positive process and let $p\in\rmatrices$ be a random projection. 
    The following are equivalent. 
    \begin{enumerate}[label = (\alph*)]
        \item $p$ reduces $(T, \psi)$. 

        \item $\glop{T}{\psi}(p)\in\prmatrices$. 

        \item There is a measurable function $f:\Omega\to (0, \infty)$ such that $\glop{T}{\psi}(p)\leq f p$ almost surely. 
    \end{enumerate}
    If, moreover, $\glop{T}{\psi}$ is bounded on $L^\infty(\Omega, \matrices)$, then the above conditions are equivalent to 
    \begin{enumerate}[label = (d)]
        \item  There is a deterministic constant $C\in (0, \infty)$ such that $\glop{T}{\psi}(p)\leq C p$ almost surely. 
    \end{enumerate}
\end{lemma}
\begin{proof}
    That (a) implies (b) is clear, because $p\in\Prmatrices$. 
    To see that (b) implies (c), notice that $\glop{T}{\psi}(p)\in\prmatrices$ implies $\proj{\glop{T}{\psi}(p)}\leq p$. 
    Therefore, since $\glop{T}{\psi}(p)\geq 0$, we have 
    \begin{equation}\label{Eqn:Red proj lemma, Eqn 1}
        \glop{T}{\psi}(p)
        \leq 
        \|\glop{T}{\psi}(p)\|_\infty \proj{\glop{T}{\psi}(p)}
        \leq 
        \|\glop{T}{\psi}(p)\|_\infty p
    \end{equation}
    almost surely, 
    so $f := \|\glop{T}{\psi}(p)\|_\infty + 1$ fulfills the requirements of (c).\footnote{We add 1 to $\|\glop{T}{\psi}(p)\|_\infty$ because we require that $f > 0$ almost surely.}
    To see that (c) implies (a), note that, because $\Prmatrices$ is spanned by its positive elements, it suffices to show that any positive $x\in \Prmatrices$ fulfills $\glop{T}{\psi}(x)\in \Prmatrices$.
    Note that such $x$ satisfies
    \begin{equation*}
        0  
        \leq 
        x
        \leq 
        \|x\|_\infty \proj{x}
        \leq 
        \|x\|_\infty p. 
    \end{equation*}
    So, since $\glop{T}{\psi}$ is a positive linear map and since $\psi\in\positivesuperops$ almost surely, we have that 
    \begin{equation*}
        0\leq \glop{T}{\psi}(x)
        \leq 
        \glop{T}{\psi}( \|x\|_\infty p)
        =
        \koopman{T}(\|x\|_\infty)
        \glop{T}{\psi}(p)
        \leq 
        \koopman{T}(\|x\|_\infty)
        f p,
    \end{equation*}
    where the last inequality holds by (c). 
    It follows that $\glop{T}{\psi}(x)\in\prmatrices$.
    Now assume that $\glop{T}{\psi}$ is bounded on $\Linftymatrices$.
    Then it follows that there is deterministic $C\in (0, \infty)$ such that $\|\glop{T}{\psi}(x)\|_\infty \leq C \|x\|_\infty$ almost surely for all $x\in\rmatrices$. 
    So, assuming (b), by inspecting the proof that (b) implies (c), it is clear that we may take $f = C$, so we see that (b) implies (d). 
    Conversely, it is clear that (d) implies (c), which concludes the proof. 
\end{proof}
\begin{lemma}\label{Lem:L is reduced by P iff L dagger is reduced by I - P}
    Let $(T, \psi)$ be a stationary positive process and let $p\in\rmatrices$ be a random projection. 
    Then $p$ reduces $(T, \psi)$ if and only if $\mbI - p$ reduces $(T, \psi)^\dagger$.
\end{lemma}
\begin{proof}
    Assume $p\in\RedProj{(T, \psi)}$. 
    Then $\glop{T}{\psi}(p)(\mbI - p) = 0$, hence 
    \begin{equation*}
        0 
            =
        \mbE_\mu\Big[
            \inner{
                \glop{T}{\psi}(p)
            }{
            \mbI - p
            }
        \Big]
            =
        \mbE_\mu\Big[
            \inner{
                p
            }{
            \glopdag{T}{\psi}(\mbI - p)
            }
        \Big],
    \end{equation*}
    which implies that $p\glopdag{T}{\psi}(\mbI - p)p= 0$ almost surely. 
    Since $\glopdag{T}{\psi}(\mbI - p)\geq 0$, by, e.g., \cite[Proposition 1.3.2]{Bhatia2009PositiveMatrices}, we conclude
    \begin{equation*}
        \glopdag{T}{\psi}(\mbI - p)
        =
        (\mbI - p) \big[\mfM_{T, \psi}^\dagger(\mbI - p)\big] (\mbI - p),
    \end{equation*}
    so by Lemma \ref{Lem:Equivalent characterizations of reducibility} we see $\mbI - p$ reduces $\glopdag{T}{\psi}$, i.e., $\mbI - p$ reduces $(T^{-1}, \koopman{T}(\psi)) = (T, \psi)^\dagger$.
    The argument is concluded by symmetry. 
\end{proof}
\begin{cor}\label{Cor:L irred iff L dagger irred}
    A stationary positive process $(T, \psi)$ is irreducible if and only if $(T, \psi)^\dagger$ is irreducible. 
\end{cor}
We find multiple occasions to apply the following simple lemma. 
\begin{lemma}\label{Lem:Positive fixed points give reducing projections for L}
    For a stationary positive process $(T, \psi)$, if $z\in\rpositives$ satisfies $\glop{T}{\psi}(z) = z$, then $\proj{z}$ reduces $(T, \psi)$. 
\end{lemma}
\begin{proof}
    Let $E = \set{z = 0}$, and define $\lambda_{\min}, \lambda_{\max}:\Omega\to (0, \infty)$ by
    \begin{equation*}
        \text{$\lambda_{\min} = 1_{\Omega\setminus E}\min\seq{\sigma(z)\cap\mbR^{>0}} + 1_E$ and $\lambda_{\max} = 1_{\Omega\setminus E}\max\sigma(z)$} + 1_E.
    \end{equation*}
    Then $0 \leq \lambda_{\min}\proj{z}\leq z\leq \lambda_{\max}\proj{z}$ by construction. 
    So, the positivity of $\glop{T}{\psi}$ gives 
    \begin{equation*}
        0
            \leq 
        \koopman{T}(\lambda_{\min})\glop{T}{\psi}(\proj{z})
            \leq 
        \glop{T}{\psi}(z)
            =
        z
            \leq 
        \lambda_{\max}\proj{z}.
    \end{equation*}
    Hence,
    \begin{equation*}
        \glop{T}{\psi}(\proj{z})
           \leq 
           \cfrac{\lambda_{\max}}{\koopman{T}(\lambda_{\min})} \proj{z}.
    \end{equation*}
    By Lemma \ref{Lem:Equivalent characterizations of reducibility} and the fact that $\lambda_{\max}, \lambda_{\min}>0$, the result is proved.
\end{proof}
We now direct our attention to the question of the existence of minimal nontrivial reducing projections, restricting our attention to ergodic positive processes $(\theta, \psi)$ such that with probability $1$, $\psi(\rho)\neq 0$ for any state $\rho$, equivalently,
\begin{equation}\label{eq:faithfulness}
    \mu\bigbracket{\ker(\psi)\cap\positives = \set{0}} = 1 \ .
\end{equation}
We begin with the observation that any reducing projection $p\in\rmatrices$ of an ergodic positive processes satisfying the faithfulness condition \eqref{eq:faithfulness} necessarily satisfies  $\mu[p = 0]\in\set{0, 1}$. 
\begin{lemma}\label{Lem:Nonzero reducing projections are almost surely nonzero}
Let $(\theta, \psi)$ be an ergodic positive process such that  $\mu\bigbracket{\ker(\psi)\cap\positives = \set{0}} = 1$, and let $p\in\rmatrices$ be a random projection. 
If $p$ reduces $(\theta, \psi)$, then 
\begin{equation*}
    \mu\bigbracket{
    \tr{p} = \operatorname{rank}(p)\geq 1
    }
    \in\set{0, 1}.
\end{equation*}
That is, $\mu[p = 0]\in\set{0, 1}$ for all $p\in\RedProj{(\theta, \psi)}$
\end{lemma}
\begin{proof}
    If $p\in\RedProj{(\theta, \psi)}$, Lemma \ref{Lem:Equivalent characterizations of reducibility} implies there is measurable $f:\Omega\to(0, \infty)$ such that $\glop{\theta}{\psi}(p)\leq f p$.
    In particular, if $\omega\in\Omega$ is such that $p_\omega = 0$, then $\glop{\theta}{\psi}(p)_\omega = 0$. Thus $1_E\glop{\theta}{\psi}(p) = 0$, with $E=\{p=0\}$
    By the faithfulness condition on $\psi$, this implies that $1_E\koopman{\theta}(p) = 0$, from which we conclude that $\set{p = 0}\setminus \theta^{-1}(\set{p = 0})$ is a null set.
    The ergodicity of $\theta$ implies $\mu[p = 0]\in\set{0,1}$. 
\end{proof}
Now, as noted in the introduction, recall that $\RedProj{(T, \psi)}$ is partially ordered by $\leq$, i.e., $p\leq q$ if and only if $p_\omega\leq q_\omega$ for almost every $\omega\in\Omega$. 
\begin{definition}[Order minimal reducing projections]\label{Def:Order minimal projections}
    Let $(T, \psi)$ be a stationary positive process.
    We call $p\in\RedProj{(T, \psi)}$ an order minimal reducing projection for $(T, \psi)$ if for all $q\in\RedProj{(T, \psi)}$, $q\leq p$ implies $p = q$. 
\end{definition}
Recall that $\RedProj{(T, \psi)}$ consists of only nonzero elements, therefore we only call $p$ order minimal if it is nonzero. 
Our first order of business is demonstrating the existence of order minimal reducing projections. 
\begin{prop}\label{Prop:There are minimal reducing projections for L}
    Let $(\theta, \psi)$ be an ergodic positive process such that $\mu\bigbracket{\ker(\psi)\cap\positives = \set{0}} = 1$. If $q\in\RedProj{\theta,\psi}$ is an order reducing projection, then there is a minimial order reducing projection $p$ for $(\theta,\psi)$ such that $p\le q$. In particular, since $\mbI \in \RedProj{\theta,\psi}$ there is an order minimal reducing projection for $(\theta, \psi)$. 
\end{prop}
\begin{proof}
     We apply Zorn's lemma with some caution to ensure measurability.  Let $\RedProj{}(q)$ be the set of order reducing projections $p\in \RedProj{\theta,\psi}$ with $p\le q$.  Note that $q\in\RedProj{}(q)$, so this set is non-empty.
     Let $\mcP = \{p_i\}_{i\in I}$ be a totally ordered decreasing set in $\RedProj{}(q)$, i.e., $\mcP$ is an ordered set, $p_i\le q$, and $i\geq j$ implies $p_i\leq p_j$. 
     It is tempting to take $p = \inf_{i\in I}p_i$, which is in some sense order minimal, but if $I$ is uncountable, $p$ need not be measurable, so some care is required in constructing the minimal element of $\mcP$. 
     Towards this end, for all $i\in I$ and $r = 1, \dots, d$, let 
     \begin{equation*}
         E_i^{(r)}
          := 
        \Big\{
            \omega\in\Omega
            \,:\, 
            \tr{p_{i; \omega}}\geq r
        \Big\}\in\mcF.
     \end{equation*}
     Since $\mcP$ is totally ordered, for any $i, j\in I$, $p_i\leq p_j$ if and only if $\tr{p_j}\leq \tr{p_j}$ almost surely. 
     So, $p_i\leq p_j$ if and only if $E_i^{(r)}\subseteq E_j^{(r)}$ for all $r = 1, \dots, d$. 
     Now, for each $r = 1, \dots, d$, define $m_r := \inf_i \mu\left[E_{i}^{(r)}\right].$
    For each $n\in\mbN$, define $i_n\in I$ recursively by $i_{n+1} > i_{n}$ and 
    \begin{equation*}
        \mu\left[E_{i_{n+1}}^{(r)}\right]\leq m_r + \frac{1}{n+1}
    \end{equation*}
    for all $r\in\{1, \dots, d\}$. 
    Then if we take $S^{(r)} := \displaystyle\bigcap_n E_{i_n}^{(r)}$, it is clear that 
    \begin{equation*}
        \mu\bigbracket{S^{(r)}} = \lim_n \mu\left[E_{i_n}^{(r)}\right] = m_r,
    \end{equation*}
    since $E_{i_{n+1}}^{(r)}\subseteq E_{i_{n}}^{(r)}$ almost surely for all $n\in\mbN$. 
     We now define $p\in\rmatrices$ to be the random projection $p = \inf_n p_{i_n}$, which is measurable as the infimum is taken over a countable set, and we claim that $p\leq p_i$ for all $i\in I$.
      To see this, fix $i\in I$, and suppose there is $N\in\mbN$ such that $i_N\geq i$.
    Then it holds that $E_{i_N}^{(r)}\subseteq E_{i}^{(r)}$ almost surely for all $r$, hence $p_{i_N}\leq p_i$ almost surely. 
    Because $p\leq p_{i_N}$, it holds that $p\leq p_i$. 
    On the other hand, if $i$ is such that $i > i_n$ for all $n$, then $p_i\leq p_{i_n}$ for all $n$ hence $p_i\leq p$. 
    Also, $E_i^{(r)}\subseteq S^{(r)}$ almost surely for all $r\in \{1, \dots, d\}$. 
    By the definition of $m_r$ and the construction of $S^{(r)}$, it follows that 
    \begin{equation*}
        \mu\bigbracket{S^{(r)}}
        \geq 
        \mu\left[E_i^{(r)}\right]
        \geq 
        m_r = \mu\bigbracket{S^{(r)}}.
    \end{equation*}
    Thus, $\mu\left[S^{(r)}\triangle E_i^{(r)}\right] = 0$ for all $r$. 
    So, $E_i^{(r)} = S^{(r)}$ almost surely for all $r$, hence $\tr{p_i}= \tr{p}$ almost surely, whereby we conclude $p \leq p_i$ so that $p = p_i$. 
    I.E., $p$ is a lower bound on $\mcP$ with respect to $\leq$. 
    Clearly $p\le q$. We claim that $p\in\RedProj{(\theta, \psi)}$, which by Zorn's lemma allows us to conclude the existence of minimal reducing projections in $\RedProj{}(q)$ and completes the proof. 
    To see this, it is clear that $p\neq 0$, since by Lemma \ref{Lem:Nonzero reducing projections are almost surely nonzero} we have that 
    \begin{equation*}
        \tr{p}
        =
        \inf_n
        \tr{p_{i_n}}
        \geq 1
    \end{equation*}
    almost everywhere.
    So, all that remains is to show that $p$ is a reducing projection. 
    For this, we note that
    \begin{equation*}
        \glop{\theta}{\psi}(p)
        \leq 
        \glop{\theta}{\psi}(p_{i_n})
        \leq 
        f_{i_n} p_{i_n}
    \end{equation*}
    holds for all $n$ for some $f_{i_n}:\Omega\to(0, \infty)$, by Lemma \ref{Lem:Equivalent characterizations of reducibility}.
    In particular, $\proj{\glop{\theta}{\psi}(p)}\leq p_{i_n}$ for all $n$. 
    By the infimum property of $p$, therefore, we conclude that $\proj{\glop{\theta}{\psi}(p)}\leq p$, so $p\in\RedProj{(\theta, \psi)}$.
    %
\end{proof}
\begin{remark}
    If $T:\Omega\to\Omega$ is not ergodic, there are not necessarily \textit{nonzero} minimal reducing projections for $\mfM_{T, \psi}$. 
    This is clear by considering the specific case that $d = 1$, $\phi = \operatorname{Id}$, and $T = \operatorname{Id}$, where the underlying probability space is $[0, 1]$ with the Lebesgue measure on Borel sets. 
    Indeed, in this case, $\rmatrices = \mathbb{M}_1(\Omega)$ is just the set of measurable functions $\Omega\to\mbC$ and the projections are precisely indicator functions $1_E$ for $E\subset [0, 1]$ Borel. 
    Moreover, all projections reduce $(T, \phi)$ trivially. 
    But the totally ordered set $\{1_{[0, 1/n]}\}_{n\in\mbN}$ of (reducing) projections has no nonzero lower bound, which shows there are no nonzero minimal reducing projections for $(T, \phi)$.
\end{remark}
There is another notion of reducibility one finds in the literature, of a more dynamical nature, defined as follows. 
\begin{definition}[Dynamically minimal reducing projections]\label{Def:Dynamical minimal projections}
    Let $(T, \psi)$ be a stationary positive process. 
    We call $p\in\RedProj{(T, \psi)}$ a dynamically minimal reducing projection for $(T, \psi)$ if for all sequences $\seq{f_n:\Omega\to\mbR^{>0}}_{n\geq 0}$ of measurable functions and all nonzero and positive $x\in\Prmatrices$, we have 
    \begin{equation*}
        \sup_{n}\proj{f_0 x + f_1 \glop{T}{\psi}(x) + \cdots + f_n \glop{T}{\psi}^n(x)}
        =
        p
    \end{equation*}
    almost everywhere. 
\end{definition}
Note that since the set of projections in $\matrices$ is an order lattice, the supremum 
\begin{equation*}
    \sup_{n}\proj{f_0 x + f_1 \mfM_{T, \psi}(x) + \cdots + f_n \mfM_{T, \psi}^n(x)}
\end{equation*}
in the above definition always exists almost everywhere. 
We begin by making the observation that the above supremum is independent of $\seq{f_n:\Omega\to\mbR^{>0}}_{n\geq 0}$.
\begin{lemma}\label{Lem:Char_of_dyn_minimality}
    Let $(T, \psi)$ be a stationary positive process, and fix a sequence  $(f_n:\Omega\to\mbR^{>0})_{n\in\mbN}$ of strictly positive measurable functions. 
    Then for all positive $x\in\rmatrices$, the random projection
    \begin{equation*}
        \sup_{n}\proj{f_0 x + f_1 \mfM_{T, \psi}(x) + \cdots + f_n \mfM_{T, \psi}^n(x)}
    \end{equation*}
    is independent of choice of $(f_n:\Omega\to\mbR^{>0})_{n\in\mbN}$. 
    In particular, $p\in\RedProj{(T, \psi)}$ is dynamically minimal if and only if there exists a sequence $(g_n:\Omega\to\mbR^{>0})_{n\in\mbN}$ of measurable functions such that 
    \begin{equation*}
        \sup_{n}\proj{g_0 x + g_1 \mfM_{T, \psi}(x) + \cdots + g_n \mfM_{T, \psi}^n(x)}
        =
        p
    \end{equation*}
    holds almost surely for all nonzero positive $x\in\Prmatrices$.
\end{lemma}
\begin{proof}
    This follows from the fact that for any set $\{x_0, \dots, x_n\}$ of positive matrices and any tuples $(c_0, \dots, c_n)$ and $(c_0', \dots, c_n')$ of constants such that $c_j, c_j'\in (0, \infty)$ for all $j$, it holds that 
    \begin{equation*}
        \proj{c_0x_0 + \cdots c_n x_n}
        =
        \proj{c_0'x_0 + \cdots c_n'x_n}. \qedhere
    \end{equation*}
\end{proof}
Next, we show that dynamical minimality is equivalent to order minimality.
\begin{prop}\label{Prop:Dyn min iff ord min}
    Let $(T, \psi)$ be a stationary positive process. Then $p\in\RedProj{(T, \psi)}$ is dynamically minimal if and only if $p$ is order minimal. 
\end{prop}
\begin{proof}
    Fix a sequence $\seq{f_n:\Omega\to\mbR^{>0}}_{n\geq 0}$ of measurable functions throughout. 
    If $p$ is not order minimal, there is $q\in\scrP_{T, \psi}$ with $q\leq p$ and $q\neq p$. 
    It is then clear that 
    \begin{equation*}
        \proj{f_0 q + f_1 \glop{T}{\psi}(q) + \cdots + f_n\glop{T}{\psi}^n(q)}\leq q\leq p
    \end{equation*}
    for all $n$ almost everywhere.
    Hence because $q\in\Prmatrices$ is positive, nonzero, and $q\neq p$, we have that $p$ is not dynamically minimal. 
    Conversely, assume $p$ is order minimal. 
    Then let $x\in\Prmatrices$ be positive and nonzero, and let $q\in\rmatrices$ be the random projection 
    \begin{equation*}
        q := 
        \sup_n
        \proj{
        f_0 x 
        +
        f_1
        \glop{T}{\psi}(x)
        +
        \cdots 
        +
        f_n \glop{T}{\psi}^n(x)
        }.
    \end{equation*}
    Since $x\in\Prmatrices$, $p$ reduces $(T, \psi)$ and $q\leq p$. 
    Moreover, $x\neq 0$, so $q\neq 0$. 
    Now all that remains to show the dynamical minimality of $p$ is that $q = p$. 
    To do this, we note that $q$ is independent of choice of $\seq{f_n:\Omega\to\mbR^{>0}}_{n\geq 0}$ by Lemma \ref{Lem:Char_of_dyn_minimality}.
    Thus, it suffices to take $f_n:\Omega\to\mbR^{>0}$ defined by 
      \begin{equation*}
        f_n 
        =
        \frac{1}{2^{n+1}}
        \seq{
        1_{\set{\mfM_{T, \psi}^n(x)= 0}}
        +
        \cfrac{
        1_{\set{\mfM_{T, \psi}^n(x)\neq 0}}
          }
          {
        \tr{\mfM_{T, \psi}^{n}(x)}
        }
           }
        .
    \end{equation*}
This sequence of strictly positive functions has the property that the sum $z := \sum_{n=0}^\infty f_n\glop{T}{\psi}^n(x)$ is a well-defined nonzero positive element of $\rmatrices$ satisfying $\proj{z} = q$ and  
\begin{equation*}
    \glop{T}{\psi}(z)
    =
    \sum_{n=0}^\infty 
    \koopman{T}(f_n)\glop{T}{\psi}^{n+1}(x).
\end{equation*}
Since $z$ is positive, if we define $\lambda_{\min}:\Omega\to\mbR^{>0}$ by $\lambda_{\min} = 1_{\set{z = 0}} + 1_{\set{z\neq 0}}\min\seq{\sigma(z)\cap\mbR^{>0}}$, then $\lambda_{\min}\proj{z}\leq z$, so by the positivity of $\glop{T}{\psi}$ we have that 
\begin{align*}
    \koopman{T}(\lambda_{\min})\glop{T}{\psi}(\proj{z})
        \leq 
    \glop{T}{\psi}(z)
        &= 
    \sum_{n=0}^\infty 
    \koopman{T}(f_n)\glop{T}{\psi}^{n+1}(x)\\
        &\leq 
    x + 
     \sum_{n=0}^\infty 
    \koopman{T}(f_n)\glop{T}{\psi}^{n+1}(x).
\end{align*}
Thus, if we define $\tilde{\lambda}_{\max}:\Omega\to\mbR^{>0}$ by $\tilde{\lambda}_{\max} = 1_{\set{\tilde{z} = 0}} + 1_{\set{\tilde{z}\neq 0}}\max\seq{\sigma(\tilde{z})\cap\mbR^{>0}}$
where $\tilde{z} = x + 
     \sum_{n=0}^\infty 
    \koopman{T}(f_n)\glop{T}{\psi}^{n+1}(x)$, 
by Lemma \ref{Lem:Char_of_dyn_minimality} we conclude that $\tilde{z}\leq \tilde{\lambda}_{\max}\proj{\tilde{z}} = \tilde{\lambda}_{\max}\proj{z}$. 
In particular, 
\begin{equation*}
    \glop{T}{\psi}(\proj{z})\leq  \cfrac{\tilde{\lambda}_{\max}}{\koopman{T}(\lambda_{\min})}\proj{z},
\end{equation*}
so by Lemma \ref{Lem:Equivalent characterizations of reducibility}, $\proj{z}$ reduces $(T, \psi)$. 
The fact that $\proj{z} = q \leq p$ and $q\neq 0$ together with the order minimality of $p$ implies that $\proj{z} = p$, so that $p$ is dynamically minimal. 
\end{proof}
By the above lemma, we may say without ambiguity that $p\in\RedProj{(T, \psi)}$ is minimal to refer to its order or dynamical minimality, and we take this convention throughout the remainder of this work. 
We now give a probabilistic characterization of minimality. 
\begin{cor}\label{Cor:Dynamical minimality, probabilistic version}
    %
    Let $(T, \psi)$ be a stationary positive process, and fix $p\in\RedProj{(T, \psi)}$. 
    Then $p$ is minimal if and only if for all sequences $\seq{f_n:\Omega\to\mbR^{>0}}_{n\geq 0}$ of measurable functions, we have
    \begin{equation}\label{Eqn:Dynamical minimality, probabilistic}
        \lim_{n\to\infty}\mu\bigbracket{
        \proj{f_0 x + f_1 \mfM_{T, \psi}(x) + \cdots + f_n \mfM_{T, \psi}^n(x)}
        = 
        p
        }
        = 1
    \end{equation}
    for all nonzero positive $x\in\Prmatrices$. 
\end{cor}
\begin{proof}
By Proposition \ref{Prop:Dyn min iff ord min}, it suffices to show that (\ref{Eqn:Dynamical minimality, probabilistic}) is equivalent to dynamical minimality. 
So, assume $p\in\RedProj{(T, \psi)}$ is dynamically minimal, fix $x\in\Prmatrices$ positive and nonzero, and let $f_n:\Omega\to\mbR^{>0}$ be a strictly positive sequence of measurable functions.
Let 
\begin{equation*}
    p_n 
        = 
    \proj{f_0x + f_1\glop{T}{\psi}(x) + \cdots + f_n\glop{T}{\psi}^n(x)}.
\end{equation*}
Then we have that $p_n\leq p_{n+1}$ for all $n\geq 0$, and, by dynamical minimality, $\sup_n p_n  = p$, so $\sup_n\tr{p_n} = \tr{p}$. 
Therefore, because $\tr{p}\in\set{1, \dots, d}$, for almost every $\omega\in\Omega$ there is $N_\omega\in\mbN$ such that $n\geq N_\omega$ implies $\tr{p_{n; \omega}} = \tr{p_\omega}$. 
Because $p_n$ and $p$ are projections with $p_n\leq p$, this implies $p_{n;\omega} = p_\omega$ for all $n\geq N_\omega$. 
So, we have just shown that $\mu[\omega\in\Omega\,\,:\,\, n\geq N_\omega]\leq \mu[p_n = p]$. 
But clearly $\lim_n\mu[\omega\in\Omega\,\,:\,\, n\geq N_\omega] = 1$, so we conclude (\ref{Eqn:Dynamical minimality, probabilistic}) holds. 
    Conversely, suppose that  (\ref{Eqn:Dynamical minimality, probabilistic}) holds. 
    Then, as argued above, we know that $1 = \lim_n \mu[p_n = p] = \mu\brackets{\bigcup_n\set{p_n = p}}$.
    So, since $p_n\leq p$, $\sup_n p_n = p$, which concludes the proof. 
\end{proof}
\begin{remark}
    For $p\in\RedProj{(T, \psi)}$ minimal, there need not be $N\in\mbN$ for which 
    \begin{equation*}
        \mu\bigbracket{
         p = 
         \proj{f_0 x + f_1 \mfM_{T, \psi}(x)+
            \cdots  + f_N\mfM_{T, \psi}^N(x)}
        }=1,
    \end{equation*}
    even under the assumption that $(T, \psi)$ is an ergodic quantum process. 
    Indeed, this is already witnessed at the level of dynamical systems: 
    consider the case of $d = 1$,
    $\Omega = [0, 1)$, $T:\Omega\to\Omega$ given by $T(s) = s + r \mod[0, 1)$ for some irrational number $r\in\mbR\setminus\mbQ$, and $\mu$ the Lebesgue measure. 
    Then $T$ is ergodic and invariant for $\mu$, 
    and if we consider $\psi:\mbM_1\to\mbM_1$ the identity map, we have that $\mfM_{T, \psi}$ is simply the map $\koopman{T}$ on the set of measurable maps $\Omega\to\mbC$. 
    By ergodicity of $T$, $\mfM_{T, \psi}$ is irreducible, so $P = 1$ is a dynamically minimal reducing projection. 
    But if we consider a Cantor set $C$ with $\mu(C) = 1/2$ and let $x = 1_C$, then for all $n\in\mbN$, we have that 
    \begin{equation*}
        \proj{x
        +
        \mfM_{T, \psi}(x)
        +
        \cdots 
        +
        \mfM_{T, \psi}^n(x)}
        =
        1_{\bigcup_{k=0}^nT^{-k}(C)},
    \end{equation*}
    and, because $[0, 1)\setminus T^{-k}(C)$ is a dense open set for all $k$, we have that
    \begin{equation*}
        [0, 1)\setminus \seq{\bigcup_{k=0}^nT^{-k}(C)}
            =
        \bigcap_{k=0}^n[0, 1)\setminus T^{-k}(C)
    \end{equation*}
    is a dense open set for all $n$, so is in particular a nonempty open set.
    Thus, $\mu\brackets{{\bigcup_{k=0}^nT^{-k}(C)}}\in (0, 1)$ for all $n$. 
    In total, we have shown that there is nonzero $x\geq 0$ such that 
    \begin{equation*}
         \mu\bigbracket{
         p = 
         \proj{x + \mfM_{T, \psi}(x)+
            \cdots  + \mfM_{T, \psi}^n(x)}
        } < 1
    \end{equation*}
    for all $n$, which is what we wanted to demonstrate.
\end{remark}
\subsection{Existence and uniqueness of stationary states}\label{Subsec:Operator L}
We now study the fixed points of $\glop{T}{\psi}$, beginning with some notation and terminology. 
\begin{definition}[Stationary states and fixed points of $(T, \psi)$]
    Let $(T, \psi)$ be a superoperator process. 
    We say that $x\in\rmatrices$ is stationary under $(T, \psi)$ if $\glop{T}{\psi}(x) = x$. 
    To denote the set of all $x\in\rmatrices$ stationary under $(T, \psi)$, we write $\FixedPts{(T, \psi)}$. 
    We say $\ArbRho\in\rmatrices$ is a stationary state for $(T, \psi)$ if $\ArbRho\in\FixedPts{(T, \psi)}\cap\rstates$. 
\end{definition}
The following theorem of Beck and Schwartz is our starting point for studying $\FixedPts{(T, \psi)}$.
We recount this theorem in a slightly different form than is given in \cite{Beck1957ATheorem}
\begin{theorem*}[Beck \& Schwartz, 1957 \cite{Beck1957ATheorem}]\label{Thm:Beck and Schwartz}
    Let $\scrX$ be a reflexive Banach space and let $(S, \Sigma, m)$ be a probability space. 
    Let $s\mapsto B_s$ be a strongly measurable\footnote{I.E., for all $x\in \scrX$, $s\mapsto B_s(x)$ is measurable when $\scrX$ is given the Borel $\sigma$-algebra induced by its norm.} function taking values in $\scrB(\scrX)$. 
    Suppose that $\|B_s\|\leq 1$ for almost every $s\in S$. 
    Let $h$ be a measure-preserving transformation in $(S, \Sigma, m)$. 
    Then for each $X\in L^1(S, \scrX)$, there is $\bar{X}\in L^1(S, \scrX)$ such that 
    \begin{equation*}
        \lim_{M\to\infty}
        \frac{1}{M}
        \sum_{N=1}^M
        B_s\cdots B_{h^{N-1}(s)}
            \big(X_{h^N(s)}\big)
        =
        \bar{X}_s
    \end{equation*}
    strongly in $\scrX$ almost everywhere in $S$, and $\bar{X}_s = B_s\big(\bar{X}_{h(s)}\big)$ almost everywhere in $S$. 
    Moreover, $\bar{X}$ is also the limit in $L^1(S, \scrX)$ sense. 
\end{theorem*}
Note that, for a superoperator process $(T, \psi)$, $\mfM_{T, \psi}\in\bops{L^1(\Omega, \matrices)}$ whenever $\mu\bigbracket{|||\psi|||\leq 1} = 1$ for some norm $|||\cdot|||$ on $\superops$ induced by a norm $|||\cdot|||$ on $\matrices$.
Because $\superops$ is a finite-dimensional Banach space, strong measurability of $\omega\mapsto\psi_\omega$ is assured. 
Therefore, taken with the Uniform Boundedness Principle \cite[Theorem III.14.1]{Conway2007AAnalysis}, the above theorem of Beck and Schwartz gives the following. 
\begin{prop}\label{App:Prop:Cesaro means exist for L, L1 version}
Let $(T, \psi)$ be a superoperator process. 
If $\mu\bigbracket{|||\psi|||\leq 1} = 1$ for some complete norm $|||\cdot|||$ on $\matrices$, then for all $x\in\Lonematrices$ the limit 
\begin{equation*}
 \mfE_{T, \psi}(x)
 :=
     \lim_{N\to\infty}
    \frac{1}{N}
    \sum_{n=1}^N
    \mfM_{T, \psi}^n(x)
\end{equation*}
exists in $\Lonematrices$ and almost surely.
Moreover, $\mfE_{T, \psi}$ defines a positive map in $\bops{L^1(\Omega, \matrices)}$ with the property that $\mfE_{T, \psi}\mfM_{T, \psi} = \mfM_{T, \psi}\mfE_{T, \psi} = \mfE_{T, \psi} = \mfE^2_{T, \psi}$, where $\mfM_{T, \psi}$ is viewed as an element of $\bops{L^1(\Omega, \matrices)}$. 
\end{prop}
Under the assumption that $\psi$ is trace preserving, this may be used to conclude the following. 
\begin{lemma}
    Let $(T, \psi)$ be a stationary positive trace preserving process. 
    Let $\mcI$ denote the $\sigma$-algebra generated by sets $E\in\mcF$ with $T^{-1}(E) = E$. 
    Then for all $x\in\Lonematrices$,
    \begin{equation*}
        \tr{\mfE_{T, \psi}(x)}
            =
        \mbE_\mu[\tr{x}\,\vert\,\mcI]
    \end{equation*}
    almost surely. 
    In particular, there exists a stationary state for $\mfM_{T, \psi}$.
\end{lemma}
\begin{proof}
    Because $\psi$ is trace preserving, 
    \begin{equation*}
        \tr{\mfE_{T, \psi}(x)}
            =
        \lim_{N\to\infty}
        \frac{1}{N}
        \sum_{n=1}^N
        \tr{\koopman{T^n}(x)}
    \end{equation*}
    holds almost everywhere. 
    Thus, from the Birkhoff Ergodic Theorem \cite[Theorem 1.14]{Walters1982AnTheory}, we have that 
    \begin{equation*}
        \tr{\mfE_{T, \psi}(x)}
            =
        \mbE[\tr{x}\,\vert\,\mcI].
    \end{equation*}
    To conclude the last part, simply consider $z = \mfE_{T, \psi}(\mbI)$. 
    Then what we have just shown implies that $\tr{z} = \mbE[\tr{\mbI}\,\,\vert\,\,\mcI] = d$, so because $\mfM_{T, \psi}(z) = z$ by construction, we see that $\uniqrho := d^{-1}z$ satisfies $\uniqrho\in\rstates$ and $\mfM_{T, \psi}(\varrho) = \uniqrho$, which ends the proof. 
\end{proof}
By a similar argument, we can establish a slight improvement of the previous result. 
Note that if $\psi$ is almost surely trace preserving, then $\mu\bigbracket{\ker(\psi)\cap\mbP_d = \set{0}} = 1$. 
\begin{lemma}\label{Lem:Stationary states for min red proj's}
    Let $(\theta, \phi)$ be an ergodic positive trace preserving process.
    Let $p\in\RedProj{(\theta, \phi)}$ be minimal. 
    Then there is $\uniqrho\in\rstates$ stationary under $(\theta, \phi)$ such that $\proj{\uniqrho} = p$. 
\end{lemma}
\begin{proof}
    Let $z = \mfE_{\theta, \phi}(p)$. 
    Then $\glop{\theta}{\phi}(z) = z$, and, since $p$ reduces $(\theta, \phi)$, we know that $z\in\Prmatrices$. 
    Moreover, by Lemma \ref{Lem:Positive fixed points give reducing projections for L}, $\proj{Z}$ reduces $\mfM_{\theta, \phi}$, hence $\proj{z}\leq p$. 
    On the other hand, by the previous lemma taken with Lemma \ref{Lem:Nonzero reducing projections are almost surely nonzero}, the ergodicity of $\theta$ implies that $\mcI$ is a trivial $\sigma$-algebra, hence 
    \begin{equation*}
        \tr{z}
            =
        \mbE[\tr{p}]
            \geq 
        1.
    \end{equation*}
    So, since $z\geq 0$, we necessarily have that $z$ is nonzero.
    Thus, $\proj{z}$ is nonzero, so by the minimality of $p$, $\proj{z} = p$. 
    Now consider $\varrho := \tr{z}^{-1}z$. 
    Then, because $\tr{z} = \mbE[\tr{z}]$, it holds that $\koopman{\theta}(\tr{z}) = \tr{z}$, hence
    \begin{equation*}
        \mfM_{T, \phi}(\varrho)
        =
        \varrho.
    \end{equation*}
    Because $\varrho\in\rstates$ by construction, the proof is done. 
\end{proof}
We have thus established relevant basic existence results. 
We now concern ourselves with the question of which uniqueness properties are associated to minimal nontrivial reducing projections. 
The following result is our main technical lemma towards this end. 
\begin{lemma}\label{Lem:Self-adjoint fixed points are signed}
    Let $(\theta, \phi)$ be an ergodic positive trace preserving process.
    Let $p\in\RedProj{(\theta, \phi)}$ be minimal, and assume $x\in\FixedPts{(\theta, \phi)}\cap\Prmatrices$ is self-adjoint. 
    Then either $x\geq 0$ or $-x\geq 0$, i.e., either $x$ is positive or $-x$ is positive. 
\end{lemma}
\begin{proof}
    We start with the basic observation that if $x\in\FixedPts{(\theta, \phi)}$ and $x_{\theta(\omega)}\ge 0$, then, because $\phi$ is positive, $0\leq \phi_{\theta(\omega)}(x_{\theta(\omega)}) = x_{\omega}$. 
    Thus, $\theta^{-1}\set{x\geq 0}\subseteq \set{x\geq 0}$, so $\mu[x\geq 0]\in\set{0, 1}$ since $\theta$ is ergodic. Similarly, we have $\mu[-x\geq 0]\in\set{0, 1}$. 

    To complete the proof of the lemma, it suffices to show that if $\mu[-x\ge 0] =0$, then $\mu[x\ge 0]=1$. 
    Given a self-adjoint $z\in \Prmatrices$, let $\lambda_1^\downarrow(z)\ge \lambda_2^\downarrow(t)\cdots \ge \lambda_{\tr{p}}^\downarrow(z)$ denote the eigenvalues of $z$ restricted to the range of $p$, written with multiplicity in non-increasing order.
    Thus our aim is to show that if $\lambda_1^\downarrow(x) >0$ almost surely, then $\lambda_{\tr{p}}^\downarrow(x)\ge 0$ almost surely.  

    Suppose that $x \in \FixedPts{(\theta, \phi)}$ is such that $\lambda_1^\downarrow(x) >0$ almost surely.  We will show that $\lambda_{\tr{p}}^\downarrow(x)\ge 0$ almost surely by interpolating between $x$ and $\uniqrho$. To this end, define $z(\cdot):[0, \infty)\to\FixedPts{(\theta, \phi)}$ by 
$$    
        z(t) := x + t \uniqrho ,
$$   
    Since the projection onto $\ran z(t)$ is reducing and $p$ is minimal, we have $\ran z(t)=\ran p$ or $z(t)=0$, where the latter case holds if and only $x = -t\uniqrho\le 0$ which contradicts our assumption that $\lambda_1^\downarrow(x) >0$. It follows that
    \begin{equation}\mu\left [\lambda_{j}^\downarrow(z(t))=0 \text{ for some } j=1,\ldots,\tr{p} \right ]=0 . \label{eq:no zero eigenvalues}\end{equation}
    
    Let $f(t)=\mu[\lambda^\downarrow_{\tr{p}}(z(t))<0]$. We will show that $f(0)=0$ by proving that $f$ is constant and $f(t)\to 0$ as $t\to \infty$. First, we note that $f(t)\in \{0,1\}$ since $1-f(t)=\mu[z(t)\ge 0]$ and $z(t)\in \FixedPts{(\theta, \phi)}\cap\Prmatrices$.  Thus to prove that $f$ is constant it suffices to prove continuity.  Because $0\le \uniqrho\le 1$, it follows that $t\mapsto\lambda^\downarrow_{\tr{p}}(z(t))$ is almost surely non-decreasing and continuous, so that
    \begin{equation}\label{eq:right continuity}\set{\lambda^\downarrow_{\tr{p}}(z(t))<0} = \bigcup_{s>t} \set{ \lambda^\downarrow_{\tr{p}}(z(s))<0 } . 
    \end{equation}
    Thus $f(t)$ is continuous from the right by monotone convergence.  To prove continuity from the left, we note that  $\mu[\lambda_{\tr{p}}(z(t))=0]=0$, by eq.\ \eqref{eq:no zero eigenvalues}, which implies that $f(t)=1-\mu[\lambda^\downarrow_{\tr{p}}(z(t))>0]$. By an argument similar to that used to obtain eq.\ \eqref{eq:right continuity}, we find that
    $$ \set{\lambda^\downarrow_{\tr{p}}(z(t))>0} = \bigcup_{s<t} \set{ \lambda^\downarrow_{\tr{p}}(z(s))>0 } .  $$
    By monotone convergence, $f(t)$ is continuous from the left, and thus constant, equal to either $0$ or $1$. 
    
    It remains to compute the limit as $t\to \infty$, which is essentially continuity from the left at $\infty$. Note that $\lambda_{\tr{p}}^\downarrow (z(t)) > 0$ for $t> - \lambda_{\tr{p}}^\downarrow(x)/\lambda_{\tr{p}}^\downarrow(\uniqrho)$ by Weyl's perturbation inequality for self adjoint matrices \cite[Corollary III.2.6]{Bhatia2013MatrixAnalysis}. Thus
    $$\set{\lambda^\downarrow_{\tr{p}}(\uniqrho) >0 } \subseteq \bigcup_{s>0} \set{\lambda^\downarrow_{\tr{p}}(z(s))> 0 } ,$$
    from which we conclude that $f(t) \to 0$ as $t\to \infty$ by monotone convergence.  Thus $f(0)=0$ as was to be shown.
\end{proof}
\begin{cor}\label{Cor:Minimal reducing projections have nondegenerate fixed point space}
Let $(\theta, \phi)$ be an ergodic positive trace preserving process.
For minimal $p\in\RedProj{(\theta, \phi)}$, there is $\uniqrho\in\rstates$ with $\proj{\uniqrho} = p$ such that the $\mbC$-vector space $ \PFixedPts{(\theta, \phi)}
        :=
    \Prmatrices\cap\FixedPts{(\theta, \phi)}$
    is spanned by $\varrho$. 
\end{cor}
\begin{proof}
    Let $\uniqrho\in\rstates$ be as in Lemma \ref{Lem:Stationary states for min red proj's}, and let $x\in\PFixedPts{(\theta, \phi)}$ be nonzero.
    Because $\phi$ is positive, both $2^{-1}(x + x^*)$ and $(2i)^{-1}(x - x^*)$ are stationary under $(\theta, \phi)$.
    So, to show $\PFixedPts{(\theta, \phi)} = \mbC\uniqrho$, we may assume without loss of generality that $x$ is self-adjoint, and, by Lemma \ref{Lem:Self-adjoint fixed points are signed}, we may further assume that $x \geq 0$. 
    Also, since $\mfE_{\theta, \phi}(x) = x$, we have that $\tr{x} = \mbE[\tr{x}]>0$, so by replacing $x$ with $\tr{x}^{-1}x$, we may assume, in fact, that $x\in\rstates$. 
    In particular, $\varrho, x\in \rstates\cap\Prmatrices$. 
    Now consider $z= x - \uniqrho$. Then $z=z^*$ is stationary under $(\theta,\phi)$, so   by Lemma \ref{Lem:Self-adjoint fixed points are signed}, either $z\geq 0$ or $-z\geq 0$.  Since $\tr{z}=0$, we conclude that $z=0$, i.e., $x=\uniqrho$.
\end{proof}
\subsubsection{Proof of Theorem \ref{Thm:Minimal projections}}\label{Subsec:Proof of minimal reducing theorem}
We are now ready to prove Theorem \ref{Thm:Minimal projections}. Recall the statement of the theorem. 
\Minimalprojectionstheorem*
\begin{proof}
We prove the theorem for ergodic positive trace preserving processes $(\theta, \phi)$, i.e., we do not require complete positivity of $\phi$. 
Now, as discussed in Remark \ref{Rmk:Relationship between operator and process}, the definition of reducing projections given in the introduction and those given in the technical body above differ. 
Nevertheless, all of our technical results apply under the general requirement that $\theta$ is measure preserving and ergodic. 
So, if we let $T = \theta^{-1}$ and let $\psi = \koopman{\theta}(\phi)$, and notice that a random projection $p\in\rmatrices$ reduces $(\theta, \phi)$ as defined in the introduction if and only if $p$ reduces $(T, \psi)$ as defined in Definition \ref{Def:Reducing projections}, then we see that proving the theorem requires us to prove the analogous statements about $(T, \psi)$, which is what we do now.
We have already seen in Proposition \ref{Prop:There are minimal reducing projections for L} that minimal reducing projections exist. 
So all that remains is to show the equivalence of (a), (b), (c), and (d). 
To do this, we show that (a), (b), and (c) are equivalent, then we show that (c) implies (d) and that (d) implies (a).
First, we know that (a) implies (c) from Corollary \ref{Cor:Minimal reducing projections have nondegenerate fixed point space}, and it is clear that (c) implies (b), as we have already seen that $\tr{p}^{-1}\mfE_{T, \psi}(p)\in\rstates$ is stationary under $(T, \psi)$. 
We now show that (b) implies (a). 
To see this, suppose that (b) holds. Then $p$ is a reducing projection by \ref{Lem:Positive fixed points give reducing projections for L}. By \ref{Prop:There are minimal reducing projections for L} there is a minimal reducing projeciton $q\le p$.   If $\uniqrho' = \mbE[\tr{q}]^{-1}\mfE_{T, \psi}(q)$, then $\uniqrho'\in\rstates$ is stationary under $(T, \psi)$ and $\proj{\varrho'} = q$. For $t\in (0,1)$, it follows that $t \uniqrho + (1-t)\uniqrho'$ is stationary and has $\proj{t\uniqrho +(1-t)\uniqrho'} = p $. Hence, by (b), 
$$
\uniqrho = t\uniqrho +(1-t)\uniqrho' 
$$
and thus $\uniqrho=\uniqrho'$, so $p=q$. Therefore $p$ is minimal.
Next, we show that (c) implies (d).
To do this, we begin by noting that (c) implies $\mfE_{T, \psi}(\arbrho) = \uniqrho$ for all $\arbrho\in p\rstates p$.
Therefore, for any essentially bounded, positive, and nonzero $x\in\Prmatrices$, Proposition \ref{App:Prop:Cesaro means exist for L, L1 version} and the fact that $\proj{\uniqrho} = p$ together imply
\begin{align*}
    0 < \inner{\varrho}{x}_{\Ltwomatrices}
    &=
    \lim_N
    \frac{1}{N}\sum_{n=1}^N
    \inner{\glop{T}{\psi}^n(\eta)}{x}_{\Ltwomatrices},
        \\
    &=
    \lim_N \frac{1}{N}\sum_{n=1}^N
    \inner{\phi^{(n)}(\arbrho)}{\koopman{\theta}^n(x)}_{\Ltwomatrices}
\end{align*}
almost surely,
where we used the fact that $\theta$ is measure preserving. 
From this, (d) is clear. 
Lastly, we show that (d) implies (a), by contrapositive. 
So, assume that (a) does not hold, and let $q\leq p$ be a minimal reducing projection for $(T, \psi)$ and let $\uniqrho\in\rstates$ be stationary under $(T, \psi)$ with $\proj{\uniqrho} = q$. 
Let $x = p - q$, and note that $x$ is positive, nonzero, and not equal to $p$. 
Moreover, 
\begin{equation*}
    \tr{\phi^{(n)}_\omega(\uniqrho_\omega) x_{\theta^n(\omega)}}
    =
    \koopman{\theta^n}
    \seq{\tr{\glop{T}{\psi}(\uniqrho)x
    }
    }_\omega 
    =
    0
\end{equation*}
for almost every $\omega\in\Omega$, since $\glop{T}{\psi}(\uniqrho)\in \rmatrices(q)$ and $q\perp p - q$. 
So, (d) does not hold, which concludes the proof. 
\end{proof}
We now briefly note the following further characterization of minimality that follows by standard ergodic theoretic arguments. 
\begin{prop}
    Let $(\theta, \phi)$ be an ergodic positive trace preserving process. 
    Then $p\in\RedProj{(\theta,\phi)}$ is minimal if and only if there exists $n\in\mbN$ such that for all $\arbrho\in\rstates$ and essentially bounded nonzero $x\in\rpositives$, if $\arbrho, x\in\Prmatrices$, then 
        \begin{equation*}
            \mu\brackets{
            \omega\in\Omega\,\,:\,\, 
            \tr{\phi^{(n)}_\omega(\arbrho_\omega)x_{\theta^n(\omega)}}
            >0
            }
            >0.
        \end{equation*}
\end{prop}
\begin{proof}
    The forward implication is clear from Theorem \ref{Thm:Minimal projections} (d), so we only need to show the reverse implication. 
    To do this, we note that 
    \begin{equation}\label{Eqn:Prop_other_class_of_min, Eqn 1}
        \mu\brackets{\bigcup_{n\geq 0}
        \set{\omega\in\Omega\,\,:\,\,
        \tr{\phi^{(n)}_\omega(\arbrho_\omega)x_{\theta^n(\omega)}}>0}
        }\in\set{0, 1}.
    \end{equation}
    Indeed, if for $\omega\in\Omega$, there exists $n\in\mbN$ such that $\tr{\phi^{(n)}_\omega(\arbrho_\omega)x_{\theta^n(\omega)}}>0$ holds for all $\arbrho, x\in\Prmatrices$ as described in the statement of the proposition, then 
    \begin{equation*}
        \tr{\phi^{(n+1)}_{\theta^{-1}(\omega)}
        (\arbrho_{\theta^{-1}(\omega)})
        x_{\theta^{n+1}(\theta^{-1}(\omega))}
        }
        =
        \tr{\phi^{(n)}_\omega\big(\phi_{\theta^{-1}(\omega)}( \arbrho_{\theta^{-1}(\omega)})\big)
        x_{\theta^n(\omega)}}>0
    \end{equation*}
    for all $\arbrho, x\in\Prmatrices$ as described in the statement of the proposition, since $p$ reduces $(\theta, \phi)$ and $\phi_{\theta^{-1}(\omega)}( \arbrho_{\theta^{-1}(\omega)})\in\rstates$. 
    Therefore, (\ref{Eqn:Prop_other_class_of_min, Eqn 1}) holds by ergodicity of $\theta$. 
    So, by Theorem \ref{Thm:Minimal projections} (d), $p$ is minimal. 
\end{proof}
\subsection{Recurrence and transience}\label{Subsec:Recurrence}
We have seen that stationary positive trace preserving processes $(T, \psi)$ have a canonical stationary state, namely $\uniqrho = d^{-1}\mfE_{T, \psi}(\mbI)$. 
Since $\uniqrho$ is a stationary state, $\proj{\uniqrho}$ is a reducing projection for $(T, \psi)$. 
The object of this section is the analysis of $\proj{\uniqrho}$.
\begin{definition}\label{Def:Recurrent and transient projections}
Let $(T, \psi)$ be a stationary positive trace preserving process, and let $\varrho = d^{-1}\mfE_{T, \psi}(I)$.
We write $\recP$ to denote $\proj{\varrho}$, and call it the recurrent projection for $(T, \psi)$.
    We write $\transP$ to denote the orthogonal complement $\mbI - \recP$ of $\recP$, and call $\transP$ the transient projection for $(T, \psi)$. 
We say $(T, \psi)$ is recurrent if $\recP = \mbI$ and we call $(T, \psi)$ dynamically ergodic if $\recP$ is a minimal reducing projection. 
\end{definition}
\begin{lemma}\label{Lem:Recurrent projection dominates fixed point projections}
Let $(\theta, \phi)$ be an ergodic positive trace preserving process.
Then $\proj{x}\leq \recP$ for all $x\in\FixedPts{(\theta, \phi)}$.
\end{lemma}
\begin{proof}
    First assume $x$ is self-adjoint. 
    Because  $x\in\FixedPts{(\theta, \phi)}$ and $\theta$ is ergodic, $\tr{x} = \mbE[\tr{x}]$ almost surely. 
    So, because $x$ is self-adjoint, we have $-\mbE[\tr{x}] \mbI\leq x\leq \mbE[\tr{x}]\mbI$. 
    Thus, the positivity of $\mfE_{\theta, \phi}$ and the stationarity of $x$ implies 
    \begin{equation*}
        -\mbE[\tr{x}]\mfE_{\theta, \phi}(\mbI)
        \leq x = \mfE_{\theta, \phi}(x)
        \leq 
        \mbE[\tr{x}]\mfE_{\theta, \phi}(\mbI). 
    \end{equation*}
    From this, we conclude $\proj{x}\leq \recP$, as desired. 
    For general $x$, we know that $x\in\FixedPts{(\theta, \phi)}$ if and only if both 
    \begin{equation*}
        \operatorname{Re}(x) = \cfrac{x + x^*}{2}\quad\text{and}\quad 
        \operatorname{Im}(x) = \cfrac{x - x^*}{2i}
    \end{equation*}
    are in $\FixedPts{(\theta, \phi)}$. 
    Because $\operatorname{Re}(x)$ and $\operatorname{Im}(x)$ are self-adjoint, we have that  $\proj{\operatorname{Re}(x)}\leq \recP$ and $\proj{\operatorname{Im}(x)}\leq \recP$ by the above.  
    So, because
    \begin{equation*}
        \ran(x)\subseteq \operatorname{span}\set{\ran(\operatorname{Re}(x)), \ran(\operatorname{Im}(x))}
        \subseteq
        \ran(\recP), 
    \end{equation*}
    we conclude that $\proj{x}\leq \recP$. 
\end{proof} 
\begin{remark}\label{Rem:Recurrent projection is the same as in Evans and Hoegh-Krohn}
    The above shows that, in the non-disordered setting, $\recP$ is precisely the projection onto what the authors of \cite{Evans1978SpectralC-Algebras} called the recurrent subspace. 
\end{remark}
The following lemma regarding minimal reducing projections is key. 
\begin{lemma}\label{Lem:Min red proj fixed for recurrent corner}
Let $(\theta, \phi)$ be an ergodic positive trace preserving process, and let $p\in\RedProj{(\theta, \phi)}$ be minimal. 
Then $\recP\glopdag{\theta}{\phi}(p)\recP = p$.
\end{lemma}
\begin{proof}
Since $p$ is minimal, $p = \proj{\uniqrho}$ where $\uniqrho$ is the stationary state $\mbE[\tr{p}]^{-1}\mfE_{\theta, \phi}(p)$. 
So, by Lemma \ref{Lem:Recurrent projection dominates fixed point projections}, $p\leq\recP$. 
Also, $\mbI - p$ reduces $\mfM_{\theta, \phi}^\dagger$, and $\|\phi^*\|_\infty = 1$ almost surely since $\phi^*$ is unital, so we have $\mfM_{\theta, \phi}^\dagger(\mbI - p)\leq \mbI - p$. 
Because $\mfM^\dagger_{\theta, \phi}(\mbI) = \mbI$, this implies $\mfM_{\theta, \phi}^\dagger(p) - p\geq 0.$
Since $p\leq \recP$, we have that $\recP\glopdag{\theta}{\phi}(p)\recP - p \geq 0$.
Moreover, $\recP\glopdag{\theta}{\phi}(p)\recP - p\in\rmatrices(\recP)$, so since $\proj{\mfE_{\theta, \phi}(\mbI)} = \recP$, we have 
\begin{align*}
\inner{\recP\mfM^\dagger_{\theta, \phi}(p)\recP - p}{\mfE_{\theta, \phi}(\mbI)}_{\Ltwomatrices}
        &= 
        \inner{\glopdag{\theta}{\phi}(p)}{\mfE_{\theta, \phi}(\mbI)}_{\Ltwomatrices}
        -
        \inner{p}{\mfE_{\theta, \phi}(\mbI)}_{\Ltwomatrices}\\
        &=
        \mbE\inner{p}{\mfE_{T, \psi}(\mbI)}
        -
        \mbE\inner{p}{\mfE_{T, \psi}(\mbI)}\\
        &=
        0.
\end{align*}
Since $\mfE_{\theta, \phi}(\mbI)$ is positive and satisfies $\proj{\mfE_{\theta, \phi}(\mbI)} = \recP$, and since $\recP\mfM^\dagger_{\theta, \phi}(p)\recP - p\in\rmatrices(\recP)$, this implies that $\recP\mfM^\dagger_{\theta, \phi}(p)\recP = p$, as desired. 
\end{proof} 
\begin{cor}\label{Cor:Both P and Pr - P reduce L}
Let $(\theta, \phi)$ be an ergodic positive trace preserving process, and let $\set{p_k}_{k=1}^n$ be a set of mutually orthogonal nonzero minimal reducing projections for $(\theta, \phi)$. 
Then if $q = \recP - \sum_{k=1}^n p_k$ is nonzero, $q$ is a reducing projection for $(\theta, \phi)$. 
\end{cor}
\begin{proof}
Let $p$ denote the sum $\sum_{k=1}^n p_k$. 
Because $\set{p_k}_{k=1}^n$ is a set of mutually orthogonal projections, we have that $p$ is a projection. 
Also, each $p_k\in\RedProj{(\theta, \phi)}$ is minimal, so $p_k = \proj{\uniqrho_k}$ for some stationary state $\uniqrho_k\in p_k\rstates p_k$. 
By Lemma \ref{Lem:Recurrent projection dominates fixed point projections}, this implies $p_k\leq\recP$, hence $p\leq\recP$. 
From this, it is clear that $\recP - p$ is a projection. 
Further, $p = \recP p \recP$, so $\recP - p = \recP(\mbI - p)\recP$. 
So, by the previous lemma, since all $p_k$ are minimal, it holds that $\recP\glopdag{\theta}{\phi}(p)\recP = p$.
Thus, 
\begin{align*}
    \inner{p\glop{\theta}{\phi}(\recP - p)p}{p}_{\Ltwomatrices}
    =
    \inner{\glop{\theta}{\phi}(\recP - p)}{p}_{\Ltwomatrices}
    &=
    \inner{\recP - p}{\glopdag{\theta}{\phi}(p)}_{\Ltwomatrices}\\
    &=
    \inner{\mbI - p}{\recP\glopdag{\theta}{\phi}(p)\recP}_{\Ltwomatrices}
    =
    0.
\end{align*}
so that $p\glop{\theta}{\phi}(\recP - p)p = 0$.
An application of \cite[Proposition 1.3.2]{Bhatia2009PositiveMatrices} then yields
\begin{align*}
    \glop{\theta}{\phi}(\recP - p)
        &=
    (\mbI - p)\glop{\theta}{\phi}(\recP - p)(\mbI - p)\\
        &=
    (\mbI - p)\recP \glop{\theta}{\phi}(\recP - p)\recP(\mbI - p)\\
        &= 
    (\recP - p)\mfM_{\theta, \phi}(\recP - p)(\recP - p)\in\rmatrices (\recP - p)
\end{align*}
    so that $\recP - p$ reduces $(\theta, \phi)$.
\end{proof}
\subsubsection{Proof of Theorem \ref{Thm:Recurrence}}
Proving Theorem \ref{Thm:Recurrence} is now just a matter of consolidating the above results. 
Recall the statement of the theorem under consideration. 
\Recurrencetheorem*
\begin{proof}
    Let $T = \theta^{-1}$ and $\psi = \koopman{\theta}(\phi)$. 
    Then as described in the proof of Theorem \ref{Thm:Minimal projections}, it suffices to prove the analogous statement for the process $(T, \psi)$. 
    Let $p\in\RedProj{(T, \psi)}$ be minimal. 
    Then by Lemma \ref{Lem:Nonzero reducing projections are almost surely nonzero} we have $\tr{p} > 0$. 
    Also, by Corollary \ref{Cor:Both P and Pr - P reduce L}, $\recP - p$ reduces $(T, \psi)$.
    If $\recP - p$ is nonzero, then we may proceed inductively using Corollary \ref{Cor:Both P and Pr - P reduce L} to produce a collection $\{p_i\}_{i=1}^n$ with $n\leq d$ of mutually orthogonal minimal reducing projections for $(T, \psi)$. 
    By their construction, it is clear that $\sum_{i=1}^n p = \recP$, as desired.

    For the last part, fix some $\arbrho:\Omega\to\states$.
    Then $\mfE_{T, \psi}(\arbrho)$ is positive and stationary under $\mfM_{T, \psi}$, so by Lemma \ref{Lem:Recurrent projection dominates fixed point projections} it holds that $\mfE_{T, \psi}(\arbrho)\in\rmatrices(\recP)$. 
    Thus, $\inner{\mfE_{T, \psi}(\arbrho)}{\transP}\equiv 0$. 
    Because $\|\transP\|_\infty \equiv 1$, we may apply the dominated convergence theorem to conclude that 
    \begin{align*}
        0
        = 
        \mbE\inner{\mfE_{T, \psi}(\arbrho) }{\transP}
        &= 
    \lim_{N\to\infty}
    \frac{1}{N}
    \sum_{n=1}^{N}
    \mbE
    \inner{\mfM_{T, \psi}^n(\arbrho)}{\transP}\\
    &= 
    \lim_{N\to\infty}
    \frac{1}{N}
    \sum_{n=1}^{N}
    \int_\Omega 
    \tr{\phi_{\theta^n(\omega)}\circ\cdots\circ\phi_{\theta(\omega)}(\arbrho_\omega) \transP_{\theta^n(\omega)}}\,\dee\mu(\omega),
    \end{align*}
    which is what we wanted. 
\end{proof}
\subsection{Dynamical ergodicity}\label{Subsec:Dyn erg}
We now investigate dynamical ergodicity more closely. 
Note by Theorem \ref{Thm:Minimal projections}, dynamical ergodicity implies the existence and uniqueness of a stationary state $\varrho$ for $\mfM_{T, \psi}$, and we use this fact freely in the following. 
We aim to prove Theorem \ref{Thm:Dynamically ergodic theorem}, which recall here for convenience. 
\Dynamicallyergodic*
First, note that by Proposition \ref{App:Prop:Cesaro means exist for L, L1 version}, if $(\theta, \phi)$ is an ergodic positive trace preserving process, we have that $\mfM_{\theta, \phi}$ defines a mean ergodic element of $\bops{L^1(\Omega, \matrices)}$ in the SOT sense, meaning that the SOT limit 
\begin{equation*}
    \lim_{N\to\infty}\frac{1}{N}
        \sum_{n=1}^N
        \mfM_{T, \psi}^n
\end{equation*}
exists. 
Thus, from \cite[Proposition 1.1]{Kummerer1979MeanW-algebras}, we may conclude the following lemma. 
\begin{lemma}\label{Lem:Fixed point spaces of L and L dagger are isomorphic}
     Let $(\theta, \phi)$ be an ergodic positive trace preserving process. 
    Then the vector spaces $\FixedPts{(\theta, \phi)}\cap\Lonematrices$ and $\FixedPts{(\theta, \phi)^\dagger}\cap \Linftymatrices$ are isomorphic. 
\end{lemma}
From here, we can prove the latter half of Theorem \ref{Thm:Dynamically ergodic theorem}. 
\begin{prop}\label{Prop:Dyn erg theorem second half}
    Let $(\theta, \phi)$ be an ergodic positive trace preserving process. 
    If $(\theta, \phi)$ is dynamically ergodic with unique stationary state $\uniqrho\in\rstates$, then any $\arbrho\in\rstates$ satisfies 
    \begin{equation*}
         \lim_{N\to\infty}
    \frac{1}{N}
    \sum_{n=1}^N
    \phi_\omega^{(n)}(\arbrho_\omega)
        =
        \mbE_\mu[\varrho]
    \end{equation*}
    almost everywhere.
    Moreover, for any $x\in \Linftymatrices$, we have that 
    \begin{equation*}
         \lim_{N\to\infty}
        \frac{1}{N}
        \sum_{n=1}^N
        \inner{\phi_{\theta^{-n}(\omega)}
        \circ\cdots\circ
        \phi_{\theta^{-1}(\omega)}
        (\arbrho_\omega)}{x_{\theta^{-n}(\omega)}}
        =
        \mbE\inner{\uniqrho}{x}
    \end{equation*}
    almost everywhere.
\end{prop}
\begin{proof}
By the previous lemma, if $(\theta, \phi)$ is dynamically ergodic, we know $\FixedPts{(\theta, \phi)^\dagger} = \mbC \mbI$.
So, for any $x\in \Linftymatrices$, Proposition \ref{App:Prop:Cesaro means exist for L, L1 version} implies that there is a deterministic constant $C_x$ depending only on $x$ such that 
\begin{equation*}
    \lim_{N\to\infty}
        \frac{1}{N}
        \sum_{n=1}^N(\glopdag{\theta}{\phi})^n(x)
    =
    C_x \mbI.
\end{equation*}
Therefore, for any $\arbrho\in\rstates$, we have that 
\begin{align}
    C_x = \inner{\arbrho}{\lim_{N\to\infty}
        \frac{1}{N}
        \sum_{n=1}^N(\glopdag{\theta}{\phi})^n(x)}
        &= 
         \lim_{N\to\infty}
        \frac{1}{N}
        \sum_{n=1}^N\inner{\arbrho}{(\glopdag{\theta}{\phi})^n(x)}
        \notag\\
        &= 
         \lim_{M\to\infty}
        \frac{1}{M}
        \sum_{N=1}^M
        \inner{\koopman{\theta^{-n}}(\phi)\circ\cdots\circ\koopman{\theta^{-1}}(\phi)(\arbrho)}{\koopman{\theta^{-n}}(x)}\label{Eqn:Dynamical erg, Eqn 1}
\end{align}
almost everywhere.
In particular, by applying the above computation when $x$ is deterministic, we find that the limit
\begin{equation*}
    \arbrho'_{\omega} := \lim_{N\to\infty}
    \frac{1}{N}
    \sum_{n=1}^N
    \phi_{\theta^{-n}(\omega)}\circ\cdots\circ\phi_{\theta^{-1}(\omega)}(\arbrho_\omega)
\end{equation*}
exists almost everywhere and is deterministic. 
In particular, 
\begin{align*}
\arbrho' &= \mbE_\mu[\arbrho']\\
    &=
    \lim_{N\to\infty}
    \frac{1}{N}
    \sum_{n=1}^N
    \mbE_\mu[\koopman{\theta^{-n}}(\phi)\circ\cdots\circ\koopman{\theta^{-1}}(\phi)(\arbrho)]\\
    &=
     \lim_{N\to\infty}
    \frac{1}{N}
    \sum_{n=1}^N
   \mbE_\mu[\mfM_{\theta, \phi}^n(\arbrho)]\\
     &=
     \mbE_\mu\left[
     \lim_{N\to\infty}
    \frac{1}{N}
    \sum_{n=1}^N
   \glop{\theta}{\phi}^n(\arbrho)
   \right]\\
   &=
    \mbE_\mu\!\left[\uniqrho\right],
\end{align*}
where $\uniqrho$ is the unique stationary state of $(\theta, \phi)$.
On the other hand, for any $x\in\Linftymatrices$, $C_x$ is deterministic, so (\ref{Eqn:Dynamical erg, Eqn 1}) gives
\begin{align*}
   C_x
    = 
\mbE_\mu[C_x] 
    =
\lim_{N\to\infty}
        \frac{1}{N}
        \sum_{n=1}^N\mbE_\mu\inner{\arbrho}{(\mfM_{\theta, \phi}^\dagger)^n(x)}
    &=
\lim_{N\to\infty}
        \frac{1}{N}
        \sum_{n=1}^N\mbE_\mu\inner{\mfM_{\theta, \phi}^n(\arbrho)}{x}\\
    &= 
        \mbE_\mu\inner{\mfE_{\theta, \phi}(\arbrho)}{x}\\
        &=
        \mbE_\mu\inner{\uniqrho}{x}
\end{align*}
for any $x\in \Linftymatrices$, which concludes the proof. 
\end{proof} 
We now concern ourselves with the first half of Theorem \ref{Thm:Dynamically ergodic theorem}.
To do this, we introduce some new notation. 
Let $\phi_{\operatorname{rec}}^*$ denote the random superoperator given by 
\begin{equation*}
    \phi_{\operatorname{rec}}^*
    =
    \operatorname{Ad}_{\recP}\circ
        \phi^*,
\end{equation*}
where $\operatorname{Ad}_{a}(m) = a m a$ for any $a, m\in\matrices$. 
We then let $\mfM_{\theta, \phi}^{\circlearrowleft \dagger}$ denote $\mfM_{\theta^{-1}, \koopman{\theta}(\phi_{\operatorname{rec}}^*)}$.
Note that by Lemma \ref{Lem:Min red proj fixed for recurrent corner}, we have that $\mfM_{\theta, \phi}^{\circlearrowleft \dagger}(p) = p$ for all minimal $p\in\RedProj{(\theta, \phi)}$. 
In particular, $p$ reduces $\mfM_{\theta, \phi}^{\circlearrowleft \dagger}$, so when we view $\mfM_{\theta, \phi}^{\circlearrowleft \dagger}$ as an element of $\bops{p L^\infty(\Omega, \matrices) p}$, it is precisely the Banach space dual of the map $\mfM_{\theta, \phi}\in \bops{p L^1(\Omega, \matrices) p}$. 
As before, we may now apply \cite[Proposition 1.1]{Kummerer1979MeanW-algebras} to conclude the following. 
\begin{lemma}
    Let $(\theta, \phi)$ be an ergodic trace preserving process, and let $p\in\RedProj{(\theta, \phi)}$ be minimal. 
     Then the vector spaces
     $\ker(p - \mfM_{\theta, \phi}\vert_{p L^1(\Omega, \matrices) p})\cap p L^1(\Omega, \matrices) p$ and $\ker(p - \mfM_{\theta, \phi}^{\circlearrowleft \dagger}\vert_{p L^\infty(\Omega, \matrices) p})\cap p L^\infty(\Omega, \matrices) p$ are isomorphic.
\end{lemma}
We may now finish the proof of Theorem \ref{Thm:Dynamically ergodic theorem}. 
\begin{proof}[Proof of Theorem \ref{Thm:Dynamically ergodic theorem}]
    We have already proved the second half of Theorem \ref{Thm:Dynamically ergodic theorem}, so it remains to show the first statement. 
    So, fix $\arbrho\in\rstates$ and essentially bounded $x\in\rmatrices$ with $p\arbrho p = \arbrho$ and $pxp = x$. 
    Let $T = \theta^{-1}$ and $\psi = \koopman{\theta}(\phi)$.
    Arguing as in the proof of Proposition \ref{Prop:Dyn erg theorem second half}, we find that, whenever $p\in\RedProj{(T, \psi))}$, there is a deterministic constant $C_x\in\mbC$ depending only on $x$ such that 
    \begin{equation*}
        \lim_{N\to\infty}
        \frac{1}{N}
        \sum_{n=1}^N
        (\mfM_{T, \psi}^{\circlearrowleft \dagger})^n(x)
        =
        C_x p.
    \end{equation*}
    So, we find that 
    \begin{align*}
        C_x 
        =
        \mbE_\mu[C_x]
        &=
        \mbE_\mu 
        \inner{\arbrho}{
         \lim_{N\to\infty}
        \frac{1}{N}
        \sum_{n=1}^N
        (\mfM_{T, \psi}^{\circlearrowleft \dagger})^n(x)
        }\\
        &= 
         \lim_{N\to\infty}
        \frac{1}{N}
        \sum_{n=1}^N
         \mbE_\mu 
        \inner{\arbrho}{
        (\mfM_{T, \psi}^{\circlearrowleft \dagger})^n(x)
        }\\
        &= 
         \lim_{N\to\infty}
        \frac{1}{N}
        \sum_{n=1}^N
         \mbE_\mu 
        \inner{\mfM_{T, \psi}^n(\arbrho)}{
        x
        }\\
        &= 
        \mbE_\mu\inner{\mfE_{T, \psi}(\arbrho)}{x}
        \\
        &= 
        \mbE_\mu\inner{\uniqrho}{x}
    \end{align*}
    where $\uniqrho$ is the unique stationary state with $\proj{\uniqrho} = p$. 
    On the other hand, since $\proj{\arbrho}\leq p \leq \recP$ hence $\recP\arbrho \recP = \arbrho$, we see that $C_x$ may be expressed as 
    \begin{align*}
        C_x
        =
        \inner{\arbrho_\omega}{ \lim_{N\to\infty}
        \frac{1}{N}
        \sum_{n=1}^N
        (\mfM_{T, \psi}^{\circlearrowleft \dagger})^n(x)_\omega}
        &= 
        \lim_{N\to\infty}
        \frac{1}{N}
        \sum_{n=1}^N
        \inner{\arbrho_\omega}{
        (\mfM_{T, \psi}^{\circlearrowleft \dagger})^n(x)_\omega}\\
        &= 
         \lim_{N\to\infty}
        \frac{1}{N}
        \sum_{n=1}^N
        \inner{
        \phi_{\theta^n(\omega)}\circ\cdots\circ\phi_{\theta(\omega)}(\arbrho_\omega)
        }{
        x_{\theta^n(\omega)}
       }
    \end{align*}
    for almost every $\omega\in\Omega$. 
    In summary, we conclude that 
    \begin{equation*}
         \lim_{N\to\infty}
        \frac{1}{N}
        \sum_{n=1}^N
        \inner{
        \phi_{\theta^n(\omega)}\circ\cdots\circ\phi_{\theta(\omega)}(\arbrho_\omega)
        }{
        x_{\theta^n(\omega)}
       }
       =
       \mbE_\mu\inner{\uniqrho}{x},
    \end{equation*}
    for almost every $\omega\in\Omega$, as desired. 
\end{proof}
\section{Proof of Theorem \ref{Thm:I.I.D. processes}}\label{Sec: IID theorem}
Let $\phi:\Xi\to\channels$ be a random quantum channel where $(\Xi, \mcG, \nu)$ is a probability space, and let $\seq{\phi_n}_{n\in\mbZ}$ be an i.i.d. sequence of random quantum channels distributed according to $\phi$. 
Then consider $\Omega = \prod_{n\in\mbZ}\Xi$ with the product $\sigma$-algebra $\Sigma$ and the probability measure $\mu = \otimes_{n\in\mbZ}\nu$ given by Kolmogorov extension. 
For all $n\in\mbZ$, let $\pi_n:\Omega\to\channels$ be the map $(\omega_i)_{i\in\mbN}\mapsto \phi_{\omega_n}$. 
We let $S:\Omega\to\Omega$ be the left shift, $(\omega_i)_{i\in\mbN}\mapsto (\omega_{i+1})_{i\in\mbN}$, which is clearly invertible and, by standard methods, ergodic. 
We then consider the ergodic quantum process $(\pi_1, S)$, and the associated operator $\glop{S^{-1}}{\pi_1}$.
We say that a projection $p\in\rmatrices$ reduces $\seq{\phi_n}_{n\in\mbZ}$ if $p$ reduces $\mfM_{S^{-1}, \pi_1}$.
We may now prove Theorem \ref{Thm:I.I.D. processes}. 
Let $\Sigma_{[-N, M]}$ denote the sub-$\sigma$-algebra of $\Sigma$ generated by functions depending only on $\prod_{i = -N}^M\Xi$ for $M, N\in \mbZ\cup\set{\pm\infty}$.
Recall the statement of the theorem. 
\IIDTheorem*
\begin{proof}
    Assume $\recP$ is deterministic and let $p$ be a minimal reducing projeciton. 
    By Lemma \ref{Lem:Min red proj fixed for recurrent corner}, we know that $\big(\operatorname{Ad}_{\recP}\circ \mfM_{S, \pi_{0}^*}\big)^n(p)  = p$
    $\mu$-almost surely. 
    Because $p$ is measurable with respect to the product $\sigma$-algebra on $\Omega = \prod_{n\in\mbZ}\Xi$, there is a sequence $x_{N}:\Omega\to\matrices$ of $\Sigma_{[-N, N]}$-measurable functions with
    \begin{equation*}
        \|p - x_{N}\| \to 0
    \end{equation*}
    almost everywhere. 
    If we let 
    \begin{equation*}
        y_N := 
          \big(\operatorname{Ad}_{\recP}\circ \mfM_{S, \pi_{0}^*}\big)^N(x_N),
    \end{equation*}
    then we see that $y_N$ is $\Sigma_{[1, \infty]}$-measurable for all $N$, since $\recP$ is deterministic. 
    Moreover, we have 
    \begin{align*}
        \|y_N - p\| 
        =
        \Big\|
            \big(\operatorname{Ad}_{\recP}\circ \mfM_{S, \pi_{0}^*}\big)^n(x_N - p)
        \Big\|_\infty
        &\leq 
        K_S^N(\|
            x_N - p
        \|)
    \end{align*}
    $\mu$-almost everywhere, since $\|\pi_0^*\|_\infty = 1$ by unitality.  Thus
    $$\mbE ( \|y_N - p\| ) \le \mbE( K_S^N(\|x_N - p\|)= \mbE(\|x_N - p\|) \to 0  $$
    by dominated convergence.
    In particular, we conclude that $p$ is $\Sigma_{[1, \infty]}$-measurable. 
    On the other hand, we know there is $\uniqrho\in\rstates$ with $\proj{\uniqrho} = p$ and $\mfM_{S^{-1}, \pi_1}^n(\uniqrho) = \uniqrho$ for all $n$. 
    Arguing as above, this implies that $\uniqrho$ is $\Sigma_{[-\infty, 0]}$-measurable. 
    In particular, because $p = \proj{\uniqrho}$ is a function of $\uniqrho$, we have that $p$  is $\Sigma_{[-\infty, 0]}$-measurable. 
    Since $p$ is both $\Sigma_{[1, \infty]}$-measurable and $\Sigma_{[-\infty, 0]}$-measurable, we discover that $p$ is independent of itself. That is, $p$ is deterministic, as desired. 
\end{proof}


%
\section*{Acknowledgments}
This material is based upon work supported by the National Science Foundation under Grant No. 2153946.
OE thanks Lubashan Pathirana for insightful discussions that took place while OE was visiting the University of Copenhagen in Summer 2024.

%

%
%

\printbibliography

@book{Conway2007AAnalysis,
    title = {{A Course in Functional Analysis}},
    year = {2007},
    author = {Conway, John B},
    edition = {second},
    publisher = {Springer New York},
    isbn = {978-1-4419-3092-7},
    doi = {https://doi.org/10.1007/978-1-4757-4383-8}
}

@article{Ekblad2025AEntanglement,
    title = {{A multiplicative ergodic theorem for bistochastic ergodic quantum processes with applications to entanglement}},
    year = {2025},
    journal = {Letters in Mathematical Physics},
    author = {Ekblad, Owen},
    number = {6},
    pages = {128},
    volume = {115},
    doi = {10.1007/s11005-025-02018-8}
}

@article{Enomoto1979AC-algebras,
    title = {{A Perron-Frobenius type theorem for positive linear maps on C*-algebras}},
    year = {1979},
    journal = {Math. Japon.},
    author = {Enomoto, Masatoshi and Watatani, Yasuo},
    number = {1},
    pages = {53--63},
    volume = {24}
}

@article{Choi1974AC-Algebras,
    title = {{A Schwarz Inequality for Positive Linear Maps on C*-Algebras}},
    year = {1974},
    journal = {Illinois J. Math.},
    author = {Choi, Main-Duen},
    number = {4},
    pages = {565--574},
    volume = {18},
    doi = {10.1215/ijm/1256051007}
}

@article{Beck1957ATheorem,
    title = {{A Vector-Valued Random Ergodic Theorem}},
    year = {1957},
    journal = {Proceedings of the American Mathematical Society},
    author = {Beck, Anatole and Schwartz, J T},
    number = {6},
    pages = {1049--1059},
    volume = {8},
    publisher = {American Mathematical Society},
    url = {http://www.jstor.org/stable/2032681},
    doi = {10.2307/2032681}
}

@article{Carbone2021AbsorptionChannels,
    title = {{Absorption in Invariant Domains for Semigroups of Quantum Channels}},
    year = {2021},
    journal = {Annales Henri Poincar{\'{e}}},
    author = {Carbone, Raffaella and Girotti, Federico},
    number = {8},
    pages = {2497--2530},
    volume = {22},
    url = {https://doi.org/10.1007/s00023-021-01016-5},
    doi = {10.1007/s00023-021-01016-5}
}

@article{Movassagh2022AnStates,
    title = {{An Ergodic Theorem for Quantum Processes with Applications to Matrix Product States}},
    year = {2022},
    journal = {Communications in Mathematical Physics},
    author = {Movassagh, Ramis and Schenker, Jeffrey},
    pages = {1175--1196},
    volume = {395},
    url = {https://doi.org/10.1007/s00220-022-04448-0},
    doi = {10.1007/s00220-022-04448-0}
}

@book{Walters1982AnTheory,
    title = {{An Introduction to Ergodic Theory}},
    year = {1982},
    booktitle = {Graduate Texts in Mathematics},
    author = {Walters, Peter},
    publisher = {Springer New York},
    url = {http://dx.doi.org/10.1007/978-1-4612-5775-2},
    isbn = {9781461257752},
    doi = {10.1007/978-1-4612-5775-2}
}

@article{Ekblad2025AsymptoticMeasurements,
    title = {{Asymptotic Purification of Quantum Trajectories under Disordered Generalized Measurements}},
    year = {2025},
    journal = {Annales Henri Poincar{\'{e}}},
    author = {Ekblad, Owen and Moreno-Nadales, Eloy and Pathirana, Lubashan and Schenker, Jeffrey},
    doi = {10.1007/s00023-025-01638-z}
}

@book{Schaefer1974BanachOperators,
    title = {{Banach Lattices and Positive Operators}},
    year = {1974},
    author = {Schaefer, H H},
    publisher = {Springer Berlin, Heidelberg},
    isbn = {978-3-642-65972-0},
    doi = {https://doi.org/10.1007/978-3-642-65970-6}
}

@article{Li2011CharacterizationsOperations,
    title = {{Characterizations of fixed points of quantum operations}},
    year = {2011},
    journal = {Journal of Mathematical Physics},
    author = {Li, Yuan},
    number = {5},
    pages = {52103},
    volume = {52},
    url = {https://doi.org/10.1063/1.3583541},
    doi = {10.1063/1.3583541}
}

@article{Pathirana2026CorrelationStates,
    title = {{Correlation Lengths for Stochastic Matrix Product States}},
    year = {2026},
    author = {Pathirana, Lubashan and Werner, Albert H}
}

@article{Zhang2024CriteriaDynamics,
    title = {{Criteria for Davies irreducibility of Markovian quantum dynamics}},
    year = {2024},
    journal = {Journal of Physics A: Mathematical and Theoretical},
    author = {Zhang, Yikang and Barthel, Thomas},
    number = {11},
    pages = {115301},
    volume = {57},
    doi = {10.1088/1751-8121/ad2a1e}
}

@article{Burgarth2013ErgodicDimensions,
    title = {{Ergodic and mixing quantum channels in finite dimensions}},
    year = {2013},
    journal = {New Journal of Physics},
    author = {Burgarth, D and Chiribella, G and Giovannetti, V and Perinotti, P and Yuasa, K},
    number = {7},
    pages = {73045},
    volume = {15},
    url = {https://iopscience.iop.org/article/10.1088/1367-2630/15/7/073045},
    doi = {10.1088/1367-2630/15/7/073045}
}

@article{Nelson2024ErgodicAlgebras,
    title = {{Ergodic quantum processes on finite von Neumann algebras}},
    year = {2024},
    journal = {Journal of Functional Analysis},
    author = {Nelson, Brent and Roon, Eric B},
    number = {4},
    pages = {110485},
    volume = {287},
    url = {https://www.sciencedirect.com/science/article/pii/S0022123624001733},
    doi = {https://doi.org/10.1016/j.jfa.2024.110485}
}

@article{Ekblad2026ErgodicMeasurements,
    title = {{Ergodic theorems for quantum trajectories under disordered generalized measurements}},
    year = {2026},
    journal = {Journal of Physics A: Mathematical and Theoretical},
    author = {Ekblad, Owen and Moreno-Nadales, Eloy and Pathirana, Lubashan},
    number = {22},
    pages = {225305},
    volume = {59},
    doi = {10.1088/1751-8121/ae7165}
}

@article{Arnold1994EvolutionaryMatrices,
    title = {{Evolutionary Formalism for Products of Positive Random Matrices}},
    year = {1994},
    journal = {The Annals of Applied Probability},
    author = {Arnold, Ludwig and Gundlach, Volker Matthias and Demetrius, Lloyd},
    number = {3},
    volume = {4},
    doi = {10.1214/aoap/1177004975}
}

@article{Roon2025FinitelyDynamics,
    title = {{Finitely Correlated States Driven by Topological Dynamics}},
    year = {2025},
    journal = {Preprint, arXiv:2507.07287},
    author = {Roon, Eric B and Schenker, Jeffrey H}
}

@article{Fannes1992FinitelyChains,
    title = {{Finitely correlated states on quantum spin chains}},
    year = {1992},
    journal = {Communications in Mathematical Physics},
    author = {Fannes, M and Nachtergaele, B and Werner, R F},
    number = {3},
    pages = {443--490},
    volume = {144},
    doi = {10.1007/BF02099178}
}

@article{Li2011FixedOperations,
    title = {{Fixed points of dual quantum operations}},
    year = {2011},
    journal = {Journal of Mathematical Analysis and Applications},
    author = {Li, Yuan},
    number = {1},
    pages = {172--179},
    volume = {382},
    url = {https://www.sciencedirect.com/science/article/pii/S0022247X11003799},
    doi = {https://doi.org/10.1016/j.jmaa.2011.04.047}
}

@article{Arias2002FixedOperations,
    title = {{Fixed points of quantum operations}},
    year = {2002},
    journal = {Journal of Mathematical Physics},
    author = {Arias, A and Gheondea, A and Gudder, S},
    number = {12},
    pages = {5872--5881},
    volume = {43},
    url = {https://doi.org/10.1063/1.1519669},
    doi = {10.1063/1.1519669}
}

@book{Kallenberg2021FoundationsProbability,
    title = {{Foundations of modern probability}},
    year = {2021},
    author = {Kallenberg, Olav and Kallenberg, Olav},
    edition = {3},
    volume = {1},
    publisher = {Springer},
    isbn = {978-3-030-61873-5},
    doi = {https://doi.org/10.1007/978-3-030-61871-1}
}

@article{Albeverio1978FrobeniusAlgebras,
    title = {{Frobenius theory for positive maps of von Neumann algebras}},
    year = {1978},
    journal = {Communications in Mathematical Physics},
    author = {Albeverio, Sergio and H{\o}egh-Krohn, Raphael},
    number = {1},
    pages = {83--94},
    volume = {64},
    url = {https://doi.org/10.1007/BF01940763},
    doi = {10.1007/BF01940763}
}

@article{Attal2006FromInteractions,
    title = {{From Repeated to Continuous Quantum Interactions}},
    year = {2006},
    journal = {Annales Henri Poincar{\'{e}}},
    author = {Attal, Stéphane and Pautrat, Yan},
    number = {1},
    pages = {59--104},
    volume = {7},
    url = {https://doi.org/10.1007/s00023-005-0242-8},
    doi = {10.1007/s00023-005-0242-8}
}

@article{Bruneau2010InfiniteDynamics,
    title = {{Infinite products of random matrices and repeated interaction dynamics}},
    year = {2010},
    journal = {Annales de l'I.H.P. Probabilit{\'{e}}s et statistiques},
    author = {Bruneau, Laurent and Joye, Alain and Merkli, Marco},
    number = {2},
    pages = {442--464},
    volume = {46},
    publisher = {Gauthier-Villars},
    url = {http://www.numdam.org/articles/10.1214/09-AIHP211/},
    doi = {10.1214/09-AIHP211}
}

@article{Farenick1996IrreducibleAlgebras,
    title = {{Irreducible Positive Linear Maps on Operator Algebras}},
    year = {1996},
    journal = {Proceedings of the American Mathematical Society},
    author = {Farenick, Douglas R},
    number = {11},
    pages = {3381--3390},
    volume = {124},
    publisher = {American Mathematical Society},
    url = {http://www.jstor.org/stable/2161316}
}

@article{Hanson2017LandauersSystems,
    title = {{Landauer’s Principle in Repeated Interaction Systems}},
    year = {2017},
    journal = {Communications in Mathematical Physics},
    author = {Hanson, Eric P and Joye, Alain and Pautrat, Yan and Raqu{\'{e}}pas, Renaud},
    number = {1},
    pages = {285--327},
    volume = {349},
    url = {https://doi.org/10.1007/s00220-016-2751-3},
    doi = {10.1007/s00220-016-2751-3}
}

@article{Pathirana2023LawProcesses,
    title = {{Law of large numbers and central limit theorem for ergodic quantum processes}},
    year = {2023},
    journal = {Journal of Mathematical Physics},
    author = {Pathirana, Lubashan and Schenker, Jeffrey},
    number = {8},
    volume = {64},
    publisher = {AIP Publishing},
    url = {http://dx.doi.org/10.1063/5.0153483},
    doi = {10.1063/5.0153483}
}

@article{Hennion1997LimitMatrices,
    title = {{Limit theorems for products of positive random matrices}},
    year = {1997},
    journal = {The Annals of Probability},
    author = {Hennion, H},
    number = {4},
    volume = {25},
    doi = {10.1214/aop/1023481103}
}

@article{Krein1948LinearSpace,
    title = {{Linear operators leaving invariant a cone in a Banach space}},
    year = {1948},
    journal = {Uspehi Matem. Nauk (N.S.)},
    author = {Kreĭn, M G and Rutman, M A},
    number = {1(23)},
    pages = {3--95},
    volume = {3},
    url = {http://mathscinet.ams.org/mathscinet-getitem?mr=27128}
}

@article{Bougron2020LinearSystems,
    title = {{Linear Response Theory and Entropic Fluctuations in Repeated Interaction Quantum Systems}},
    year = {2020},
    journal = {Journal of Statistical Physics},
    author = {Bougron, Jean-François and Bruneau, Laurent},
    number = {5},
    pages = {1636--1677},
    volume = {181},
    url = {https://doi.org/10.1007/s10955-020-02640-x},
    doi = {10.1007/s10955-020-02640-x}
}

@article{Bougron2022MarkovianSystems,
    title = {{Markovian repeated interaction quantum systems}},
    year = {2022},
    journal = {Reviews in Mathematical Physics},
    author = {Bougron, Jean-François and Joye, Alain and Pillet, Claude-Alain},
    number = {09},
    volume = {34},
    publisher = {World Scientific Pub Co Pte Ltd},
    url = {http://dx.doi.org/10.1142/s0129055x22500283},
    doi = {10.1142/s0129055x22500283}
}

@book{Neumann1932MathematischeQuantenmechanik,
    title = {{Mathematische Grundlagen der Quantenmechanik}},
    year = {1932},
    author = {Neumann, John von},
    publisher = {Springer},
    address = {Berlin}
}

@book{Bhatia2013MatrixAnalysis,
    title = {{Matrix analysis}},
    year = {2013},
    author = {Bhatia, Rajendra},
    volume = {169},
    publisher = {Springer Science {\&} Business Media},
    isbn = {978-1-4612-6857-4},
    doi = {https://doi.org/10.1007/978-1-4612-0653-8}
}

@article{Pollicott2010MaximalProducts,
    title = {{Maximal Lyapunov exponents for random matrix products}},
    year = {2010},
    journal = {Inventiones mathematicae},
    author = {Pollicott, Mark},
    number = {1},
    pages = {209--226},
    volume = {181},
    doi = {10.1007/s00222-010-0246-y}
}

@article{Kummerer1979MeanW-algebras,
    title = {{Mean ergodic semigroups on W*-algebras}},
    year = {1979},
    journal = {Acta. Sci. Math.},
    author = {K{\"{u}}mmerer, Burkhard and Nagel, Rainer},
    pages = {151--159},
    volume = {41}
}

@article{Pellegrini2009Non-MarkovianMeasurements,
    title = {{Non-Markovian quantum repeated interactions and measurements}},
    year = {2009},
    journal = {Journal of Physics A: Mathematical and Theoretical},
    author = {Pellegrini, C and Petruccione, F},
    number = {42},
    pages = {425304},
    volume = {42},
    url = {https://iopscience.iop.org/article/10.1088/1751-8113/42/42/425304},
    doi = {10.1088/1751-8113/42/42/425304}
}

@article{Carbone2020OnChannel,
    title = {{On Period, Cycles and Fixed Points of a Quantum Channel}},
    year = {2020},
    journal = {Annales Henri Poincar{\'{e}}},
    author = {Carbone, Raffaella and Jen{\v{c}}ov{\'{a}}, Anna},
    number = {1},
    pages = {155--188},
    volume = {21},
    url = {https://doi.org/10.1007/s00023-019-00861-9},
    doi = {10.1007/s00023-019-00861-9}
}

@article{Groh1983OnC-Algebras,
    title = {{On the Peripheral Spectrum of Uniformly Ergodic Positive Operators on C*-Algebras}},
    year = {1983},
    journal = {Journal of Operator Theory},
    author = {Groh, Ulrich},
    number = {1},
    pages = {31--37},
    volume = {10},
    publisher = {Theta Foundation},
    url = {http://www.jstor.org/stable/24714164}
}

@article{Grimmer2016OpenInteraction,
    title = {{Open dynamics under rapid repeated interaction}},
    year = {2016},
    journal = {Physical Review A},
    author = {Grimmer, Daniel and Layden, David and Mann, Robert B and Mart{\'{i}}n-Mart{\'{i}}nez, Eduardo},
    number = {3},
    pages = {32126},
    volume = {94},
    publisher = {American Physical Society},
    url = {https://link.aps.org/doi/10.1103/PhysRevA.94.032126},
    doi = {10.1103/PhysRevA.94.032126}
}

@article{Carbone2016OpenProperties,
    title = {{Open Quantum Random Walks: Reducibility, Period, Ergodic Properties}},
    year = {2016},
    journal = {Annales Henri Poincar{\'{e}}},
    author = {Carbone, Raffaella and Pautrat, Yan},
    number = {1},
    pages = {99--135},
    volume = {17},
    url = {https://doi.org/10.1007/s00023-015-0396-y},
    doi = {10.1007/s00023-015-0396-y}
}

@book{Eisner2015OperatorTheory,
    title = {{Operator theoretic aspects of ergodic theory}},
    year = {2015},
    author = {Eisner, Tanja and Farkas, Bálint and Haase, Markus and Nagel, Rainer},
    volume = {272},
    publisher = {Springer},
    isbn = {978-3-319-37105-4},
    doi = {https://doi.org/10.1007/978-3-319-16898-2}
}

@article{Sherman1951OrderAlgebras,
    title = {{Order in Operator Algebras}},
    year = {1951},
    journal = {American Journal of Mathematics},
    author = {Sherman, S},
    number = {1},
    pages = {227--232},
    volume = {73},
    publisher = {Johns Hopkins University Press},
    url = {http://www.jstor.org/stable/2372173}
}

@article{Kadison1951OrderOperators,
    title = {{Order Properties of Bounded Self-Adjoint Operators}},
    year = {1951},
    journal = {Proceedings of the American Mathematical Society},
    author = {Kadison, Richard V},
    number = {3},
    pages = {505--510},
    volume = {2},
    publisher = {American Mathematical Society},
    url = {http://www.jstor.org/stable/2031784}
}

@article{Ekblad2026PeriodicityProcesses,
    title = {{Periodicity in Ergodic Quantum Processes}},
    year = {2026},
    journal = {Preprint, arXiv:2604.09422},
    author = {Ekblad, Owen and Schenker, Jeffrey}
}

@incollection{Schrader2001Perron-FrobeniusIdeals,
    title = {{Perron-Frobenius Theory for Positive Maps on Trace Ideals}},
    year = {2001},
    booktitle = {Mathematical Physics in Mathematics and Physics: Quantum and Operator Algebraic Aspects},
    author = {Schrader, Robert},
    editor = {Longo, Roberto},
    pages = {361--378},
    publisher = {American Mathematical Soc.},
    address = {Providence, RI},
    doi = {https://doi.org/10.48550/arXiv.math-ph/0007020}
}

@book{Bhatia2009PositiveMatrices,
    title = {{Positive definite matrices}},
    year = {2009},
    author = {Bhatia, Rajendra},
    publisher = {Princeton university press},
    isbn = {9780691168258},
    doi = {https://doi.org/10.1515/9781400827787}
}

@article{Stormer1963PositiveAlgebras,
    title = {{Positive linear maps of operator algebras}},
    year = {1963},
    journal = {Acta Mathematica},
    author = {Stormer, Erling},
    pages = {233--278},
    volume = {110},
    doi = {10.1007/BF02391860}
}

@article{Evstigneev1974PositiveSystems,
    title = {{Positive matrix-valued cocycles over dynamical systems}},
    year = {1974},
    journal = {Uspehi Mat. Nauk},
    author = {Evstigneev, I V},
    number = {5(179)},
    pages = {219--220},
    volume = {29}
}

@article{Mierczynski2013PrincipalTheory,
    title = {{Principal Lyapunov exponents and principal Floquet spaces of positive random dynamical systems. I. General theory}},
    year = {2013},
    journal = {Transactions of the American Mathematical Society},
    author = {Mierczy{\'{n}}ski, Janusz and Shen, Wenxian},
    number = {10},
    pages = {5329--5365},
    volume = {365},
    doi = {10.1090/S0002-9947-2013-05814-X}
}

@article{Mierczynski2013PrincipalSystems,
    title = {{Principal Lyapunov exponents and principal Floquet spaces of positive random dynamical systems. II. Finite-dimensional systems}},
    year = {2013},
    journal = {Journal of Mathematical Analysis and Applications},
    author = {Mierczy{\'{n}}ski, Janusz and Shen, Wenxian},
    number = {2},
    pages = {438--458},
    volume = {404},
    doi = {10.1016/j.jmaa.2013.03.039}
}

@misc{Wolf2012QuantumTour,
    title = {{Quantum Channels and Operations - Guided Tour}},
    year = {2012},
    author = {Wolf, Michael M}
}

@article{Cusumano2022QuantumGuide,
    title = {{Quantum Collision Models: A Beginner Guide}},
    year = {2022},
    journal = {Entropy},
    author = {Cusumano, Stefano},
    number = {9},
    pages = {1258},
    volume = {24},
    publisher = {Multidisciplinary Digital Publishing Institute},
    url = {https://www.mdpi.com/1099-4300/24/9/1258},
    doi = {10.3390/e24091258}
}

@article{Ciccarello2022QuantumInteractions,
    title = {{Quantum collision models: Open system dynamics from repeated interactions}},
    year = {2022},
    journal = {Physics Reports},
    author = {Ciccarello, Francesco and Lorenzo, Salvatore and Giovannetti, Vittorio and Palma, G Massimo},
    pages = {1--70},
    volume = {954},
    url = {https://www.sciencedirect.com/science/article/pii/S0370157322000035},
    doi = {https://doi.org/10.1016/j.physrep.2022.01.001}
}

@article{Verstraete2009QuantumDissipation,
    title = {{Quantum computation and quantum-state engineering driven by dissipation}},
    year = {2009},
    journal = {Nature Physics},
    author = {Verstraete, Frank and Wolf, Michael M and Ignacio Cirac, J},
    number = {9},
    pages = {633--636},
    volume = {5},
    doi = {10.1038/nphys1342}
}

@article{Schenker2024QuenchedMeasurements,
    title = {{Quenched large deviations of Birkhoff sums along random quantum measurements}},
    year = {2024},
    journal = {arXiv preprint arxiv:2410.15465},
    author = {Schenker, Jeffrey and Raqu{\'{e}}pas, Renaud}
}

@article{Bruzda2009RandomOperations,
    title = {{Random quantum operations}},
    year = {2009},
    journal = {Physics Letters A},
    author = {Bruzda, Wojciech and Cappellini, Valerio and Sommers, Hans-Jürgen and {\.{Z}}yczkowski, Karol},
    number = {3},
    pages = {320--324},
    volume = {373},
    url = {https://www.sciencedirect.com/science/article/pii/S0375960108016885},
    doi = {https://doi.org/10.1016/j.physleta.2008.11.043}
}

@article{Bruneau2008RandomSystems,
    title = {{Random Repeated Interaction Quantum Systems}},
    year = {2008},
    journal = {Communications in Mathematical Physics},
    author = {Bruneau, Laurent and Joye, Alain and Merkli, Marco},
    number = {2},
    pages = {553--581},
    volume = {284},
    url = {https://doi.org/10.1007/s00220-008-0580-8},
    doi = {10.1007/s00220-008-0580-8}
}

@article{Nechita2012RandomStates,
    title = {{Random repeated quantum interactions and random invariant states}},
    year = {2012},
    journal = {Probability Theory and Related Fields},
    author = {Nechita, Ion and Pellegrini, Clément},
    number = {1},
    pages = {299--320},
    volume = {152},
    url = {https://doi.org/10.1007/s00440-010-0323-6},
    doi = {10.1007/s00440-010-0323-6}
}

@article{Bruneau2014RepeatedSystems,
    title = {{Repeated interactions in open quantum systems}},
    year = {2014},
    journal = {Journal of Mathematical Physics},
    author = {Bruneau, Laurent and Joye, Alain and Merkli, Marco},
    number = {7},
    pages = {75204},
    volume = {55},
    url = {https://doi.org/10.1063/1.4879240},
    doi = {10.1063/1.4879240}
}

@article{Evans1978SpectralC-Algebras,
    title = {{Spectral Properties of Positive Maps on C*-Algebras}},
    year = {1978},
    journal = {Journal of the London Mathematical Society},
    author = {Evans, David E and H{\o}egh-Krohn, Raphael},
    number = {2},
    pages = {345--355},
    volume = {s2-17},
    url = {https://londmathsoc.onlinelibrary.wiley.com/doi/abs/10.1112/jlms/s2-17.2.345},
    doi = {10.1112/jlms/s2-17.2.345}
}

@article{Movassagh2021TheoryProcesses,
    title = {{Theory of Ergodic Quantum Processes}},
    year = {2021},
    journal = {Physical Review X},
    author = {Movassagh, Ramis and Schenker, Jeffrey},
    number = {4},
    volume = {11},
    publisher = {American Physical Society (APS)},
    url = {http://dx.doi.org/10.1103/physrevx.11.041001},
    doi = {10.1103/physrevx.11.041001}
}

@book{Schaefer1999TopologicalSpaces,
    title = {{Topological vector spaces}},
    year = {1999},
    author = {Schaefer, H H and Wolff, M P},
    edition = {second},
    publisher = {Springer New York, NY},
    isbn = {978-1-4612-7155-0},
    doi = {https://doi.org/10.1007/978-1-4612-1468-7}
}

@article{Frobenius1912UberElementen,
    title = {{{\"{U}}ber Matrizen aus nicht negativen Elementen}},
    year = {1912},
    journal = {Sitzungsberichte der K{\"{o}}niglich Preu{\ss}ischen Akademie der Wissenschaften zu Berlin},
    author = {Frobenius, Georg},
    pages = {456--477},
    publisher = {K{\"{o}}nigliche Akademie der Wissenschaften Berlin}
}

@article{Groh1984UniformW-Algebras,
    title = {{Uniform Ergodic Theorems for Identity Preserving Schwarz Maps on W*-Algebras}},
    year = {1984},
    journal = {Journal of Operator Theory},
    author = {Groh, Ulrich},
    number = {2},
    pages = {395--404},
    volume = {11},
    publisher = {Theta Foundation},
    url = {http://www.jstor.org/stable/24714224}
}

@article{Groh1984UniformlyC-algebras,
    title = {{Uniformly ergodic maps on C*-algebras}},
    year = {1984},
    journal = {Israel Journal of Mathematics},
    author = {Groh, Ulrich},
    number = {2},
    pages = {227--235},
    volume = {47},
    url = {https://doi.org/10.1007/BF02760517},
    doi = {10.1007/BF02760517}
}

@article{Neumann1927WahrscheinlichkeitstheoretischerQuantenmechanik,
    title = {{Wahrscheinlichkeitstheoretischer Aufbau der Quantenmechanik}},
    year = {1927},
    journal = {Nachrichten von der Gesellschaft der Wissenschaften zu G{\"{o}}ttingen, Mathematisch-Physikalische Klasse},
    author = {Neumann, J von},
    pages = {245--272},
    volume = {1927},
    url = {http://eudml.org/doc/59230}
}

@article{Perron1907ZurMatrices,
    title = {{Zur Theorie der Matrices}},
    year = {1907},
    journal = {Mathematische Annalen},
    author = {Perron, Oskar},
    pages = {248--263},
    volume = {64},
    url = {http://eudml.org/doc/158317}
}

\end{document}